\documentclass[aps,twocolumn,superscriptaddress,groupedaddress]{revtex4}  
\usepackage{amsmath}
\usepackage{amssymb}
\usepackage{graphicx}
\usepackage{array,multirow}
\usepackage[toc,page]{appendix}
\usepackage{color}
\usepackage{ulem}

\newcolumntype{C}[1]{>{\centering\arraybackslash}m{#1}}

\newcommand{\qgsjet}{\textsc{QGSJet}}

\newcommand{\einv}{E_{\mathrm{inv}}}
\newcommand{\ecal}{E_{\mathrm{cal}}}

\begin{document}
\begin{minipage}{10cm}
\begin{flushleft}
\hspace{-0.5cm} \textbf{Published in PRD as DOI: 10.1103\slash PhysRevD.100.082003}
\end{flushleft}
\vspace{2cm}
\end{minipage}

\title{Data-driven estimation of the invisible energy of cosmic ray showers with the Pierre Auger Observatory}

\author{The Pierre Auger Collaboration}

\date{October 25th, 2019}

\begin{abstract}
\vspace{+5cm}
The determination of the primary energy of extensive air showers using the fluorescence detection technique requires an estimation of the energy carried away by particles that do not deposit all their energy in the atmosphere. This estimation is typically made using Monte Carlo simulations and thus depends on the assumed primary particle mass and on model predictions for neutrino and muon production.  In this work we present a new method to obtain the invisible energy from events detected by the Pierre Auger Observatory. The method uses measurements of the muon number at ground level, and it allows us to reduce significantly the systematic uncertainties related to the mass composition and the high energy hadronic interaction models, and consequently to improve the estimation of the energy scale of the Observatory.

\end{abstract}


\author{A.~Aab}
\affiliation{IMAPP, Radboud University Nijmegen, Nijmegen, The Netherlands}

\author{P.~Abreu}
\affiliation{Laborat\'orio de Instrumenta\c{c}\~ao e F\'\i{}sica Experimental de Part\'\i{}culas -- LIP and Instituto Superior T\'ecnico -- IST, Universidade de Lisboa -- UL, Lisboa, Portugal}

\author{M.~Aglietta}
\affiliation{Osservatorio Astrofisico di Torino (INAF), Torino, Italy}
\affiliation{INFN, Sezione di Torino, Torino, Italy}

\author{I.F.M.~Albuquerque}
\affiliation{Universidade de S\~ao Paulo, Instituto de F\'\i{}sica, S\~ao Paulo, SP, Brazil}

\author{J.M.~Albury}
\affiliation{University of Adelaide, Adelaide, S.A., Australia}

\author{I.~Allekotte}
\affiliation{Centro At\'omico Bariloche and Instituto Balseiro (CNEA-UNCuyo-CONICET), San Carlos de Bariloche, Argentina}

\author{A.~Almela}
\affiliation{Instituto de Tecnolog\'\i{}as en Detecci\'on y Astropart\'\i{}culas (CNEA, CONICET, UNSAM), Buenos Aires, Argentina}
\affiliation{Universidad Tecnol\'ogica Nacional -- Facultad Regional Buenos Aires, Buenos Aires, Argentina}

\author{J.~Alvarez Castillo}
\affiliation{Universidad Nacional Aut\'onoma de M\'exico, M\'exico, D.F., M\'exico}

\author{J.~Alvarez-Mu\~niz}
\affiliation{Instituto Galego de F\'\i{}sica de Altas Enerx\'\i{}as (I.G.F.A.E.), Universidad de Santiago de Compostela, Santiago de Compostela, Spain}

\author{G.A.~Anastasi}
\affiliation{Gran Sasso Science Institute, L'Aquila, Italy}
\affiliation{INFN Laboratori Nazionali del Gran Sasso, Assergi (L'Aquila), Italy}

% ...

\collaboration{The Pierre Auger Collaboration}
\email{auger_spokespersons@fnal.gov}
\homepage{http://www.auger.org}
\noaffiliation


\maketitle

\section{Introduction}

The flux of cosmic rays with energy above $10^{15}$ eV is so tiny that the only way to study it is to detect the
extensive air showers that primary cosmic rays produce in the Earth's atmosphere. This is typically done using arrays of particle detectors on the ground. The estimation of the shower energy is one of the most challenging problems because the conversion of the signal detected at ground level into shower energy needs detailed Monte Carlo simulations of the air showers which are subject to large uncertainties. In fact, the signal detected at ground level depends on the primary mass which is unknown, and on details of the hadronic interactions which are also unknown, because the interactions are at energies and in phase-space regions not well covered by, or not accessible to, accelerator experiments.

Above $10^{17}$ eV, the problem of energy estimation has been solved with the implementation of the fluorescence detection technique. A fluorescence telescope detects the fluorescence radiation  emitted by the nitrogen molecules of air excited by the charged particles of the shower. The fluorescence radiation is produced in proportion to the energy dissipation allowing a reconstruction of the longitudinal profile of the energy deposit ($dE/dX$) of the shower as a function of the atmospheric depth ($X$). Thus, with the fluorescence detection technique, the atmosphere is used as a calorimeter, and the integral  $\int \left(dE/dX\right) dX$  is called the calorimetric energy of the shower ($E_{\rm cal}$). $E_{\rm cal}$  underestimates the total shower energy ($E_0$) because neutrinos do not suffer electromagnetic interactions and high energy muons reach ground level after releasing only a portion of their energy into the atmosphere. Thus, an estimation of the primary energy $E_0$ with the fluorescence detection technique is obtained by adding to $E_{\rm cal}$ a correction to account for the {\it invisible} energy ($E_{\rm inv}$) carried by the particles that do not dissipate all their energy in the atmosphere. $E_{\rm inv}$, sometimes also called {\it missing} energy in the literature, was calculated for the first time by Linsley~\cite{Linsley} and it amounts to about  10\% - 20\% of the total shower energy.

The fluorescence detection technique has been successfully implemented in the Fly's Eye experiment~\cite{FlysEye} and its successor the High Resolution Fly's Eye (HiRes)~\cite{HiRes}. The Fly's Eye group used a parametrization of $E_{\rm inv}$ derived from the estimation of Linsley~\cite{Baltrusaitis}. Later, a more refined calculation of $E_{\rm inv}$  was presented in~\cite{Song}. In that work the  invisible energy was calculated using simulated showers. The method consisted of subtracting from $E_0$ the calorimetric energy calculated from the number of charged particles as a function of atmospheric depth and assuming a mean energy loss rate.

Profiting from improvements in the detail of air shower simulations, a different and better approach to calculate $E_{\rm inv}$ was presented in~\cite{Barbosa}. The strategy consisted of obtaining $E_{\rm inv}$ directly from the energy deposited in the atmosphere by the different components of air showers. This method has been used to estimate the shower energy with the fluorescence detection technique by the two largest cosmic ray observatories currently in operation: the Telescope Array (TA)~\cite{TA} and the Pierre Auger Observatory (Auger)~\cite{Auger}.

Despite the more precise calculation developed in~\cite{Barbosa}, the
invisible energy estimate is affected by the irreducible uncertainties
associated with the models describing the hadronic interactions.
While there have been some significant improvements in these models,
in particular for the first interaction at high energy for which the
inelastic cross-section and multiplicity have been severely
constrained by LHC results~\cite{Pierog-ICRC2017}, the neutrino and
muon production still suffers from the fact that the particle
identification in accelerator measurements is still poorly constrained
in the phase space relevant to air shower development.  After the
introduction of the forward $\rho^{0}$ resonance in all
models~\cite{Riehn:2015uqw}, the differences between the predictions
for the muon content have been reduced significantly, but independent
measurements still show that all models suffer from a deficit of muon
production compared with data~\cite{Dembinski-UHECR2019}.

The uncertainty in the total shower energy due to the invisible energy model is not expected to be very large because $E_{\rm inv}$ is only a modest fraction of $E_{0}$. For example,  in~\cite{Barbosa} the uncertainty in $\einv$ propagated to $E_0 $ was estimated to be about 5\%. The models to get $\einv$ can be improved further using the primary mass composition estimated with the fluorescence detectors~\cite{Auger-mass-comp}. However, we should keep in mind that the uncertainties associated with the hadronic interaction models are difficult to estimate and are ultimately unknown~\cite{Pierog-ICRC2017}.

In this paper we present a new estimate of the invisible energy obtained analizing the data collected by the Pierre Auger Observatory~\cite{Auger}. The estimation is done by exploiting the sensitivity of the water-Cherenkov detectors (WCD) of Auger to shower muons. We have developed two different analysis methods for two different ranges of zenith angles of the showers ~\cite{UHECR2016Einv,ICRC2013Einv}. In both cases, our data-driven estimation of the invisible energy allows us to reduce significantly the dependence on mass composition and hadronic interaction models.

The paper is structured in the following way. In the first section we briefly describe the Auger detectors, while the second section deals with the phenomenology of the invisible energy, addressing the basic features of air shower development that will allow us to obtain the data-driven estimation of $\einv$. Next, we describe the analysis methods, and we report the results of the estimation including the systematic uncertainties. Finally, we discuss the results and report the conclusions of this work.

\section{The Pierre Auger Observatory}
\label{Sec:Auger}

The Pierre Auger Observatory~\cite{Auger} is located in a region called Pampa Amarilla, near the town of Malarg\"ue in the province of Mendoza (Argentina), at $\sim$35$^\circ$~S latitude and an altitude of 1400 m above sea level. Auger is a {\it hybrid} observatory because the measurements are done combining the data of  a Surface Detector (SD) and a Fluorescence Detector (FD).  In this way, the tiny flux of the ultra high energy cosmic rays can be studied with the 100\% duty cycle of the SD and with the precise shower energy estimation of the FD. The calibration of the SD signals against the FD energies is done by analizing the subset of showers detected simultaneously by the two detectors, the so-called hybrid events.

The SD consists of 1660 WCDs arranged on a hexagonal grid of 1.5~km spacing extending over a total area of  $\sim 3000~{\rm km}^2$.  Each WCD unit is a plastic tank of cylindrical shape, 10~m$^2$ in area and filled to a depth of 1.2~m with purified water. The Cherenkov radiation produced in the water is detected by three photo-multiplier tubes (PMTs), each 9'' in diameter. The PMT signals are digitized by flash analog-to-digital converters (FADC) at a 40 MHz sampling rate and calibrated online continuously using the signals produced by atmospheric background muons.

The FD consists of 24 telescopes placed in four sites located along the perimeter of the Observatory that overlook the atmosphere above the surface array. Each telescope has a field of view of $30^\circ \times 30^\circ$ and is composed of a spherical mirror with a curvature radius of 3.4~m and a camera, placed on the focal surface, which has an array of 440 hexagonal pixels (22 rows $\times$ 20 columns) each equipped with a light concentrator and a PMT. The PMT signals are digitized by an ADC at a frequency of 10 MHz. The FD operates during clear and moonless nights with a duty cycle of about 14\%~\cite{AugerFD}.

The WCDs are sensitive to the electromagnetic and hadronic components of a shower. Electrons and photons are absorbed in the water and produce Cherenkov light in an amount approximately proportional to their energy, while muons produce a signal proportional to their track length. Thanks to the 1.2~m height of the WCDs, the array is also sensitive to showers arriving at large zenith angles. In these showers, the signals detected by the WCDs are dominated by muons because the electromagnetic component of the shower is largely absorbed during the long atmospheric depth traversed before reaching ground level.

Two different reconstruction techniques are used for the events recorded by the SD: one for the so-called  {\it vertical} showers with zenith angles $\theta < 60^\circ$, and one for the {\it inclined} showers with $\theta > 60^\circ$. In vertical showers, the energy estimator is $S(1000)$, the signal at $1000$ m from the core~\cite{Auger}. In inclined showers, the energy estimator is the normalization of simulated muon density maps that are used to predict the muon pattern at ground level ~\cite{HASRec}.

For the FD events, the reconstruction of the energy deposit $dE/dX$ as a function of atmospheric depth is described in~\cite{AugerFD}. It is fitted with a Gaisser-Hillas~\cite{GH,LongProf} function whose integral gives the calorimetric energy. The total shower energy is obtained by adding the invisible energy correction. The FD is also used to set the energy scale of the Observatory which is known with a systematic uncertainty of 14\%~\cite{Auger-EnSc}.  In addition, the FD measures the depth $X_{\rm max}$ at which $dE/dX$ reaches its maximum, since it is the main experimental observable used to estimate the primary mass composition~\cite{Auger-mass}.

\section{Phenomenology of the invisible energy}
\label{Sec:EinvPhenomenology}
The basic features of the development of the extensive air showers produced in the Earth's atmosphere by primary nuclei are described by the Heitler model~\cite{Heitler} and its extension to the hadronic case~\cite{Matthews}. Although simplified in several aspects, this model allows a description of the cascade which is suitable enough to serve as a guiding thread in the next sections, where the starting points of the data-driven approaches to estimate $\einv$ will be inspired by some of the expressions outlined below.

In the model, only pions are produced in the hadronic interactions (in the proportion of two charged pions for every $\pi^0$) and they all have the same energy.
The particle multiplicity ($N$) is assumed to be the same at all energies and in all interactions. The neutral pions decay almost immediately into two photons, generating an electromagnetic cascade. The charged pions interact hadronically until the average energy of the charged pions is decreased to such a level that their time-dilated decay length ($\lambda_{\rm d}$) becomes smaller than their hadronic interaction length ($\lambda_{\rm i}$).
This energy is referred to as the pion critical energy, and it is given by
\begin{equation}
\epsilon^\pi_c = \frac{E_0}{N^{n}},
\label{eq:epi}
\end{equation}
where $E_0$ is the primary particle energy and $n$ is the number of interactions suffered by the charged pions. Note that $n$ can be determined from the equation $\lambda_{\rm i} = \lambda_{\rm d} = \rho~\epsilon^\pi_c~\tau_\pi / (m_{\pi} c) $ where $\rho$ is the atmospheric density, $\tau_\pi$ is the pion lifetime and $m_{\pi}$ is its mass. As pointed out in~\cite{K-U-review}, for an isothermal atmosphere $\rho \propto \lambda_{\rm i} ~n$ and this equation becomes
\begin{equation}
n~\epsilon^\pi_c = \frac{h_{0}~m_\pi c^2}{\cos\theta ~c\tau_\pi},
\label{eq:n-epi}
\end{equation}
where $h_0$ is a constant and $\theta$ is the zenith angle of the shower. Then, combining Eq.~\eqref{eq:epi} and~\eqref{eq:n-epi}, one arrives at the interesting conclusion that both $n$ and $\epsilon^\pi_c$ depend only on $N$, $E_0$ and $\theta$, and they do not depend on the interaction length~\cite{K-U-review}. The model provides numerical values of $\epsilon^\pi_c $ of a few tens of GeV with a slow decrease with $E_0$~\cite{K-U-review}.

One important feature of the model is that the invisible energy is proportional to the number of muons $N_\mu$ reaching ground level. In fact, once the pions reach $\epsilon^\pi_c$, they decay into muons and neutrinos that are assumed to reach ground level without any interaction (the model neglects muon decay). Then the invisible energy is simply given by
\begin{equation}
\einv = \epsilon^\pi_c ~N_\mu,
\label{eq:EinvNmu}
\end{equation}
where the number of muons is equal to the number of charged pions  ($N_\mu = \left(\frac{2}{3}N\right)^n$). This expression will be the guiding thread to estimate $\einv$ with inclined showers, for which a measurement of $N_\mu$ is possible.

Another important feature of the model is that the invisible energy follows a power-law relationship with the primary energy:
\begin{equation}
\einv=\epsilon^\pi_c\left(\frac{E_0}{\epsilon^\pi_c}\right)^\beta,
\label{eq:EinvE0}
\end{equation}
where $\beta = \ln(\frac{2}{3} N) / \ln N$. Air shower simulations predict values of $\beta$ in the range from 0.88 to 0.92~\cite{K-U-review}, that correspond to values of $N$ between 30 and 200. Note that $\beta$ also fixes how the invisible energy depends on the mass number $A$ of the primary. In fact, neglecting collective effects in the first interactions so that the cascade is the superposition of $A$ cascades initiated by primary protons of energy $E_0/A$ one has:
\begin{equation}
\einv ^A=\epsilon^\pi_c\left(\frac{E_0}{\epsilon^\pi_c}\right)^\beta A^{1-\beta}.
\label{eq:EinvE0A}
\end{equation}
This is the other relationship that will be used to estimate $\einv$ from the Auger data, as it will be the guiding thread for the analysis of the vertical showers.

The Heitler model~\cite{Heitler} extended to the hadronic case~\cite{Matthews} described above provides a qualitative description of the shower cascades. For more quantitative predictions one has to use Monte Carlo simulations that take into account all the complex phenomena occurring throughout the shower development. The invisible energy of the simulated showers is calculated following the method described in~\cite{Barbosa}. Here $\einv$ is obtained by subtracting from the primary energy $E_0$ all the energy deposited into the atmosphere. The calculation counts as deposited energy the energy that would have been deposited into the atmosphere by particles whose interactions are not simulated because they have reached ground level, or because their energies are below a predefined threshold that is set to reduce the CPU time needed to simulate the shower.

This $\einv$ definition is well suited for the fluorescence reconstruction technique, given that the calorimetric energy is estimated by integrating the Gaisser-Hillas profile over all depths, including those below ground level.

\begin{figure}[]
\centering
 \centerline{
    \includegraphics[width=0.98\columnwidth,height=0.88\columnwidth] {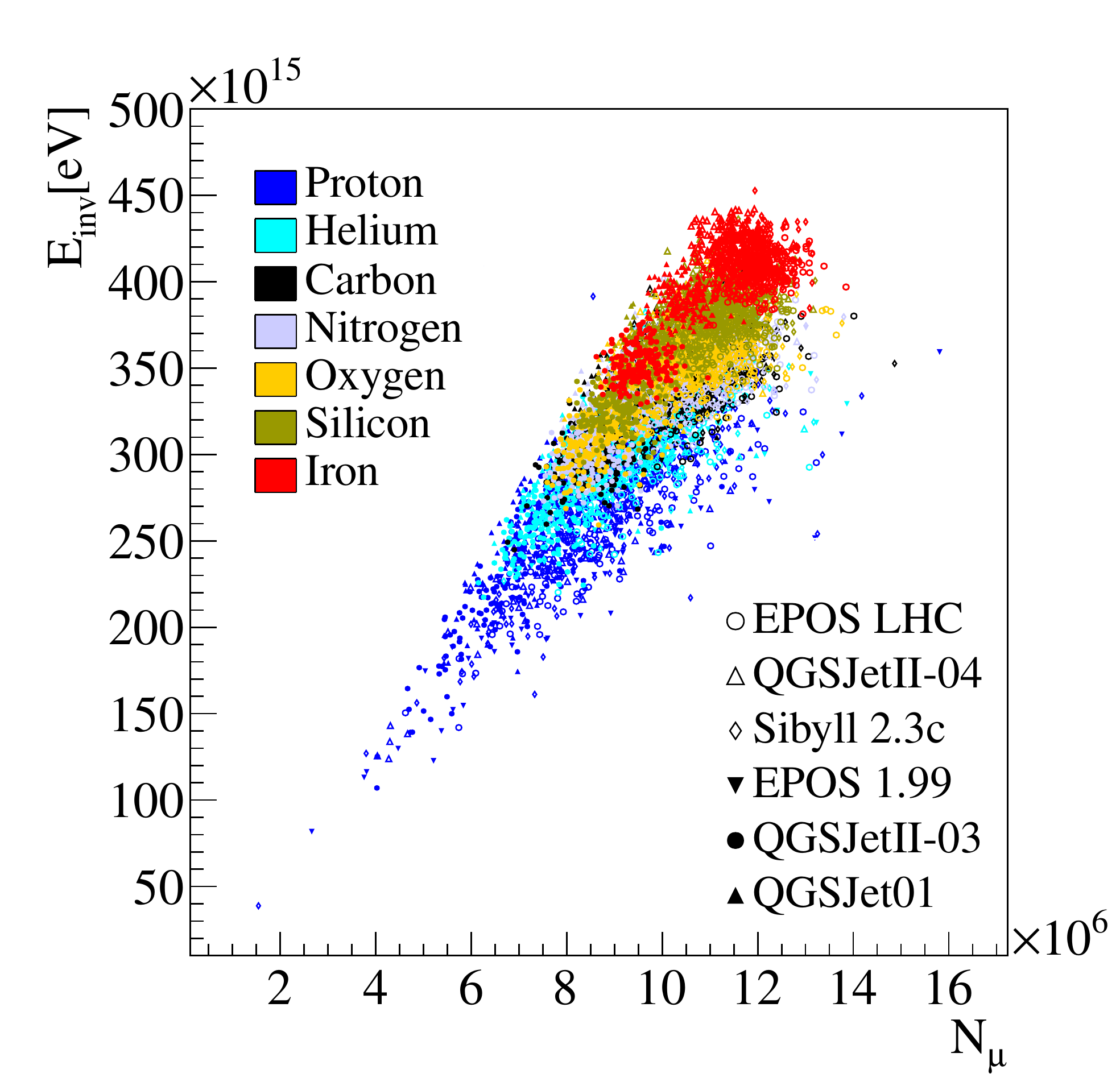}
 }
\caption{
Correlation between the invisible energy and the number of muons reaching ground level for primaries of different masses all with energy of $3 \times 10^{18}$ eV.
}
\label{fig:Einv-Nmu}
\end{figure}

\begin{figure*}[]
\centering
 \centerline{
   \includegraphics[width=1\columnwidth,height=1\columnwidth,keepaspectratio,clip] {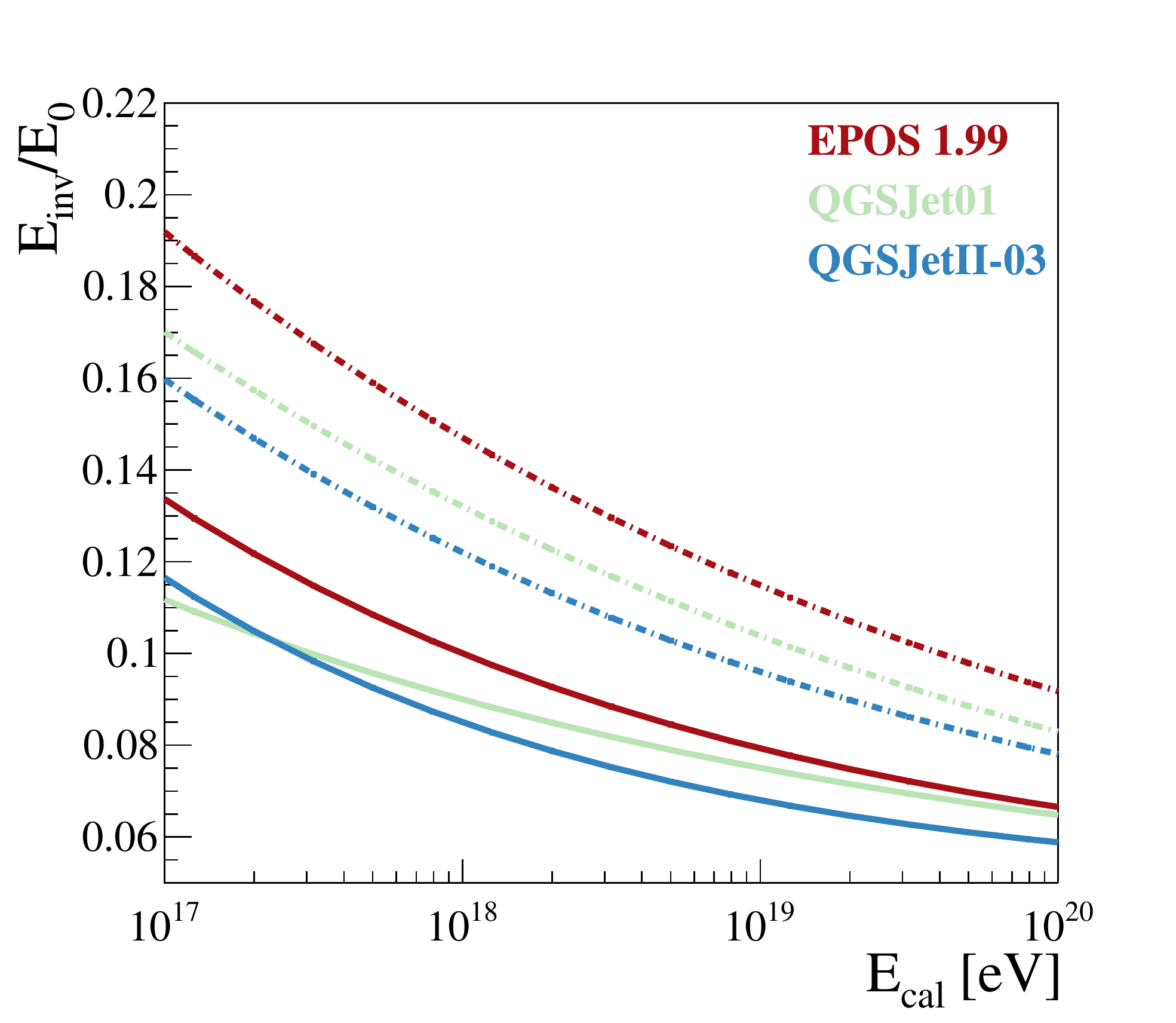}
   \includegraphics[width=1\columnwidth,height=1\columnwidth,keepaspectratio,clip] {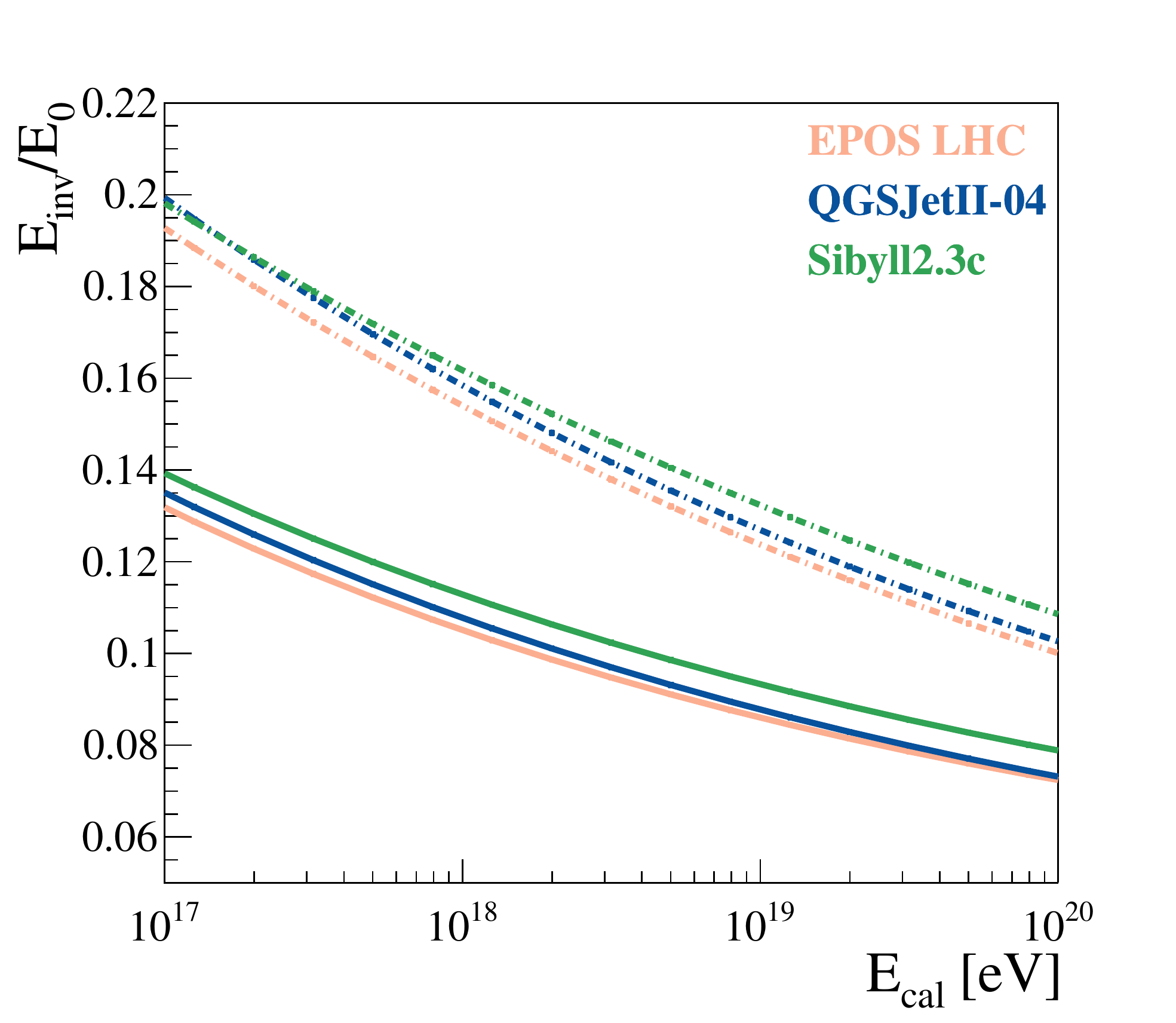}
 }
\caption{
Average invisible energy fraction as a function of the calorimetric energy calculated with Monte Carlo simulations using the hadronic interaction models tuned with LHC data (right panel) and the older models developed when the LHC data were not available (left panel). The predictions for proton and iron primaries are shown with solid and dashed lines, respectively.
}
\label{fig:Einv-MC}
\end{figure*}

The correlation between  $\einv$ and $N_\mu$ has been studied simulating showers with the CORSIKA~\cite{CORSIKA} code.
The results of the simulations for different primary masses and for the most recent hadronic interaction models
EPOS LHC~\cite{EPOS-LHC}, \qgsjet II-04~\cite{QGSJet2-04} and Sibyll2.3c~\cite{Sibyll2.3c} tuned with LHC data and the older models EPOS 1.99~\cite{EPOS}, \qgsjet II-03~\cite{QGSJet2}, \qgsjet 01~\cite{QGSJet1}, are shown in Fig.~\ref{fig:Einv-Nmu}. The simulations refer to primaries of energy $3 \times 10^{18}$ eV with a zenith angle of $60^\circ$, and $N_\mu$ is obtained by counting all muons with energy greater than 100 MeV that reach ground level at the altitude of the Observatory.

A variety of models has been used to cover different physics processes
at the origin of the muon production. According to~\cite{Matthews},
the total multiplicity and the pion charge ratio (which is linked to
the baryon and $\rho^{0}$ resonance
production~\cite{Pierog:2006qv,Ostapchenko:2013pia}) are two
fundamental parameters that drive the production of muons in air
showers. As a result, a model with a low baryon production and low
multiplicity like Sibyll 2.1~\cite{Sibyll2.1} (a version of Sibyll before the advent of the LHC data)
has the lowest muon number, while after
correction Sibyll 2.3c now has the largest muon number. QGSJET01 has a
relatively large muon production because of its high multiplicity and
despite the lack of $\rho^{0}$ resonance production. The difference
between QGSJETII-03 and QGSJETII-04 in regard to the muon production
is mostly in the $\rho^{0}$ production, while the difference between
EPOS 1.99 and EPOS LHC is mostly due to a change in the forward
baryons in high energy interactions.

In spite of the different implementations of the physics processes that lead to a very large spread in the predictions of $N_\mu$ and $\einv$, the correlation between them is good and is similar for all models and primaries considered. This suggests that it is possible to obtain a robust estimation of $\einv$ from the measurements of $N_\mu$ using the Auger inclined showers.

Other quantitative predictions of the values of $\einv$ using Monte Carlo simulations are shown in Fig.~\ref{fig:Einv-MC}. The results are presented showing the ratio of the invisible energy to the total primary energy as a function of $\ecal$. The simulations were performed using the CORSIKA~\cite{CORSIKA} code for the models
EPOS 1.99~\cite{EPOS} and \qgsjet II-03~\cite{QGSJet2} and with the AIRES~\cite{AIRES} code for \qgsjet 01~\cite{QGSJet1} (left panel). For the models tuned with the LHC data EPOS LHC~\cite{EPOS-LHC}, \qgsjet II-04~\cite{QGSJet2-04}, and Sibyll2.3c~\cite{Sibyll2.3c} we used the CONEX code~\cite{CONEX}  (right panel).

From the figure one can see the large differences in the values of $\einv$ for different primary masses and how, for a given primary mass, the spread between the predictions from different models is significantly reduced after the tuning with LHC data. Then one may argue that a precise estimation of $\einv$ can be obtained using the post-LHC models and the primary mass composition estimated from the $X_{\rm max}$ measurements~\cite{Auger-mass}. However, even after the updates with LHC data,
the models still fail to describe the muon density at ground level ~\cite{HASRec,Auger-HadInt-UHECR18, AugerMuonSize}, which can introduce unpredictable biases in the $\einv$ estimation.

Thus, the strategy followed in this paper is to estimate the invisible energy using the correlations that exist between $\einv$ and shower observables that can be measured at the Observatory, correlations that to a large extent are not sensitive to the hadronic interaction models and primary mass composition.

\section{Estimation of invisible energy using Auger data}
\label{Sec:EinvAnalysis}

The most straightforward way to estimate the invisible energy using Auger data is to use the inclined showers. In fact, for these showers it is possible to measure the total number of muons arriving at ground level which is, as seen in Sec.~\ref{Sec:EinvPhenomenology} (Eq.~\eqref{eq:EinvNmu}), an observable expected to be proportional to $E_{\rm inv}$ with a proportionality factor only marginally dependent on hadronic interaction models and primary mass.

The muon number cannot be directly measured for vertical events.  However, the invisible energy can be obtained from the energy estimator using the power-law relationship between $E_{\rm inv}$ and the total shower energy (see Eq.~\eqref{eq:EinvE0A}).

\begin{figure*}[]
\centering
 \centerline{
\includegraphics[width=1\columnwidth,height=0.9\columnwidth,clip]{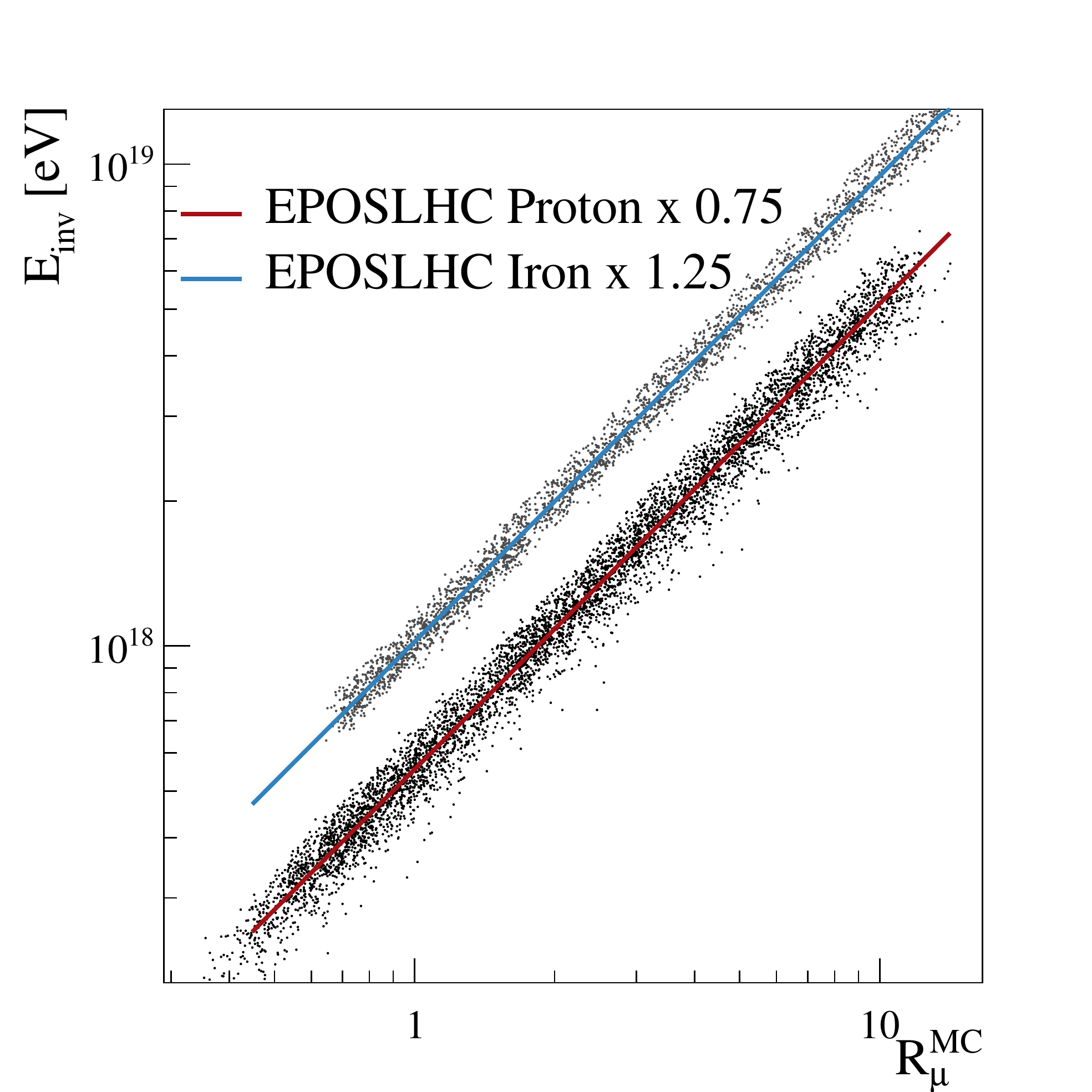}
\includegraphics[width=1\columnwidth,height=0.9\columnwidth,clip]{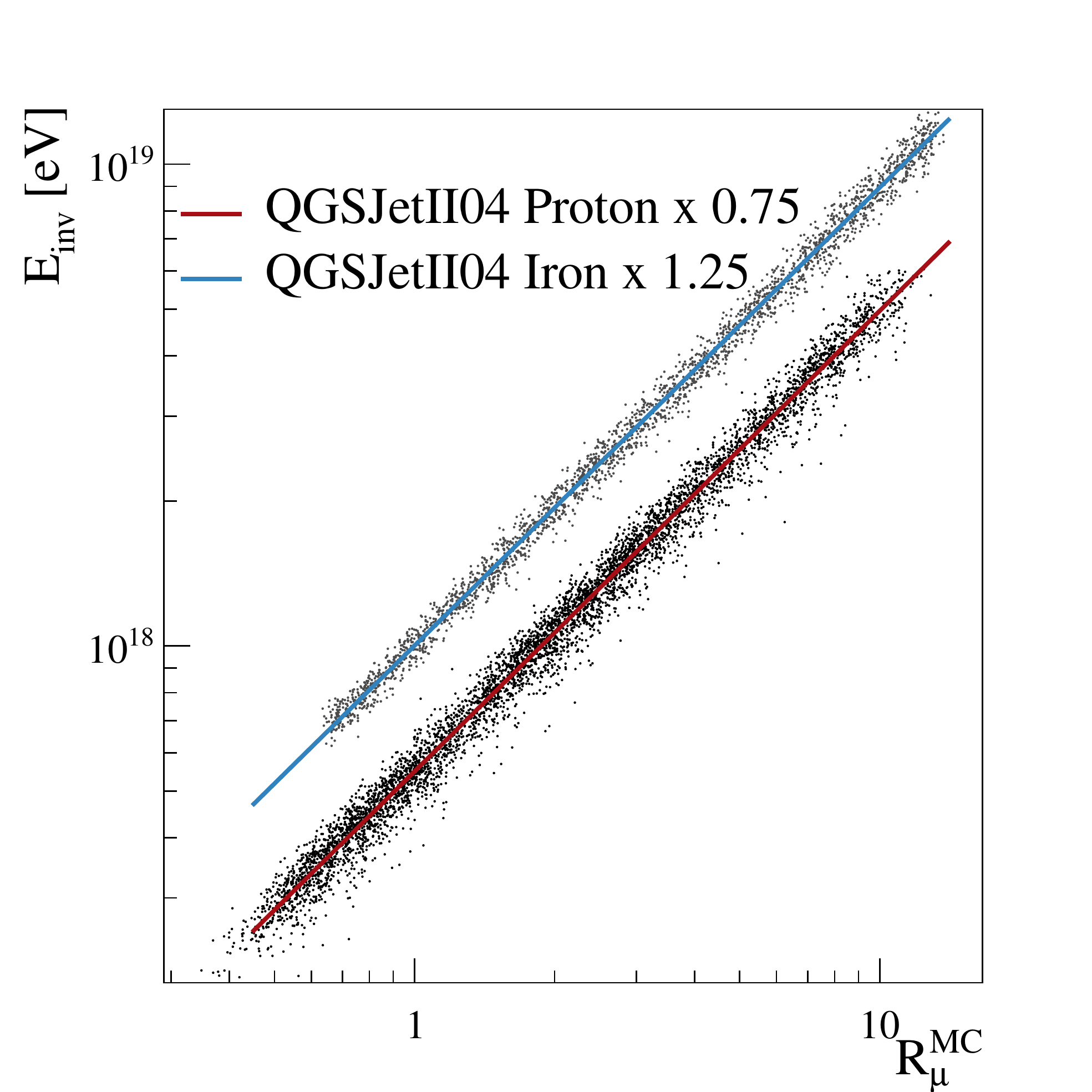}
}
\caption{Correlation between $E_{\rm inv}$ and $R_{\mu}^{\rm MC}$ for the proton (black dots) and iron (grey dots) showers simulated with the EPOS LHC~\cite{EPOS-LHC}  (left) and \qgsjet II-04~\cite{EPOS-LHC}  (right) hadronic interaction models. The red and blue lines are the results of the fit of  the power law parameters  (Eq.~\eqref{eq:EinvN19}) for protons and iron, respectively.
For a better visualization of the data, $E_{\rm inv}$ has been multiplied by the factors shown in the figure.}
\label{fig:Einv_Rmu}
\end{figure*}

\subsection{$E_{\rm inv}$ from inclined showers}
\label{Sec:Einv_Rmu}
The reconstruction of inclined events is described in~\cite{HASRec}.  The basic information used in the reconstruction is that the muon number distribution at ground level can be described by a density scaling factor that depends on the shower energy and primary mass, and by a lateral shape that, for a given  arrival direction $(\theta,\phi)$ of the shower, is consistently reproduced by different hadronic interaction models and depends only weakly on the primary energy and mass. The muon number density as a function of the position at ground $\vec{r}$ is then parametrized with
\begin{equation}
\rho_{\mu}(\vec{r}) = N_{19}~\rho_{\mu,19}(\vec{r};\theta,\phi),
\label{MDensity}
\end{equation}
where $\rho_{\mu,19}(\vec{r};\theta,\phi)$ is a reference distribution conventionally calculated
for primary protons at $10^{19}$ eV using the hadronic interaction model \qgsjet II-03~\cite{QGSJet2}, and the scale factor $N_{19}$ represents the shower size relative to the normalization of the reference distribution.

The scale factor is determined with a maximum-likelihood method based on a realistic \textsc{Geant4}~\cite{Geant4} simulation of the WCD response. The simulation is done with the Auger \mbox{$\overline{\textrm{Off}}$\hspace{.05em}\protect\raisebox{.4ex}{$\protect\underline{\textrm{line}}$}} software framework~\cite{Offline}. A residual electromagnetic signal component, mainly due to muon decays in flight, is taken into account according to model predictions~\cite{Ines-em}.

The performance of the reconstruction is validated on simulated events. For each event, the  reconstructed value of $N_{19}$ is compared with its true value $R_{\mu}^{\rm MC}$. The latter is defined as the ratio of the total number of muons at ground level to the total number of muons in the reference model. The relative deviation of $N_{19}$ from $R_{\mu}^{\rm MC}$ is within 5\% for several hadronic interaction models and primaries~\cite{AugerMuonSize}. A bias correction is then applied to $N_{19}$  in order to reduce the residuals to within 3\% of the most recent models tuned with LHC data. In this way, the corrected value of $N_{19}$, which in the following is called $R_\mu$, represents an unbiased estimator of the total number of muons at ground level.

The correlation between the invisible energy and the total number of muons at ground level is studied with Monte Carlo simulations. Two data sets are simulated: one with CORSIKA~\cite{CORSIKA} using the hadronic interaction models EPOS LHC~\cite{EPOS-LHC} and \qgsjet II-04~\cite{QGSJet2-04} and the other with AIRES~\cite{AIRES} using the model \qgsjet 01~\cite{QGSJet1}. The showers have zenith angles isotropically distributed between $60^{\circ}$ and $80^{\circ}$ and energies ranging from  ${10}^{18}$ to ${10}^{20}$ eV.

For each Monte Carlo event, we calculate the values of $E_{\rm inv}$ and of the muon number at ground level $R_{\mu}^{\rm MC}$. For all the samples of simulated events, the correlation between $E_{\rm inv}$ and $R_{\mu}^{\rm MC}$
is well described by a power-law functional form
\begin{equation}
E_{\rm inv} =  C  ~  \left(R_{\mu}^{\rm MC}\right)^{\delta},
\label{eq:EinvN19}
\end{equation}
where the values of the parameters $C$ and $\delta$ are obtained from a fit to the  events.
Examples of the correlation between $E_{\rm inv}$ and $R_{\mu}^{\rm MC}$ are shown Fig.~\ref{fig:Einv_Rmu}, where the lines show the fitted power law relationships. For all simulations, the root mean square of the distribution of residuals of the fit is less than 10\%. The values of the parameters for all the simulations are shown in table ~\ref{tab:Einv_Rmu}. All the values of $\delta$ are close to 1, showing the validity of the prediction done by the Heitler inspired model~\cite{Matthews} according to which the invisible $E_{\rm inv}$ is proportional to $N_\mu$ (see Eq.~\eqref{eq:EinvNmu}).

\begin{table}[ht!]
\caption{Values of the parameters describing the power law relationship between $E_{\rm inv}$ and $R_{\mu}^{\rm MC}$ (Eq.~\eqref{eq:EinvN19}) for proton and iron primaries and different hadronic interaction models.}
\label{tab:Einv_Rmu}
\begin{center}
\begin{tabular}{|cc|c|c|}\hline
\multicolumn{2}{|C{4.cm}|}{primary and hadronic interaction model}  & $C$ [$10^{18}$ eV] &  $\delta$ \\  \hline
             & EPOS LHC~\cite{EPOS-LHC}          & 0.739 & 0.967  \\
proton   & \qgsjet II-04~\cite{QGSJet2-04}   & 0.732 & 0.956  \\
             & \qgsjet 01~\cite{QGSJet1}              & 0.736 & 0.969   \\ \hline
             & EPOS LHC~\cite{EPOS-LHC}           & 0.816 & 0.967  \\
  iron     & \qgsjet II0-04~\cite{QGSJet2-04}  & 0.801 & 0.951  \\
              & \qgsjet 01~\cite{QGSJet1}              & 0.810 & 0.963   \\ \hline
\end{tabular}
\end{center}
\end{table}

The relationship of Eq.~\eqref{eq:EinvN19} is used to estimate the invisible energy in the data from the measurement of $R_{\mu}$ that, as seen before, is the unbiased estimator of $R_{\mu}^{\rm MC}$. Since the primary mass composition of the data is not precisely known, the estimation of the invisible energy is obtained using the parametrization of $E_{\rm inv}$ as a function of $R_{\mu}$ for a mixture of 50\% protons and 50\% iron. This is done taking the average of the two  $E_{\rm inv}$ estimations that are obtained for proton and iron primaries using the EPOS LHC~\cite{EPOS-LHC} hadronic interaction model.

\begin{figure}[]
\centering
 \centerline{
\includegraphics[width=1\columnwidth,height=0.9\columnwidth,clip]{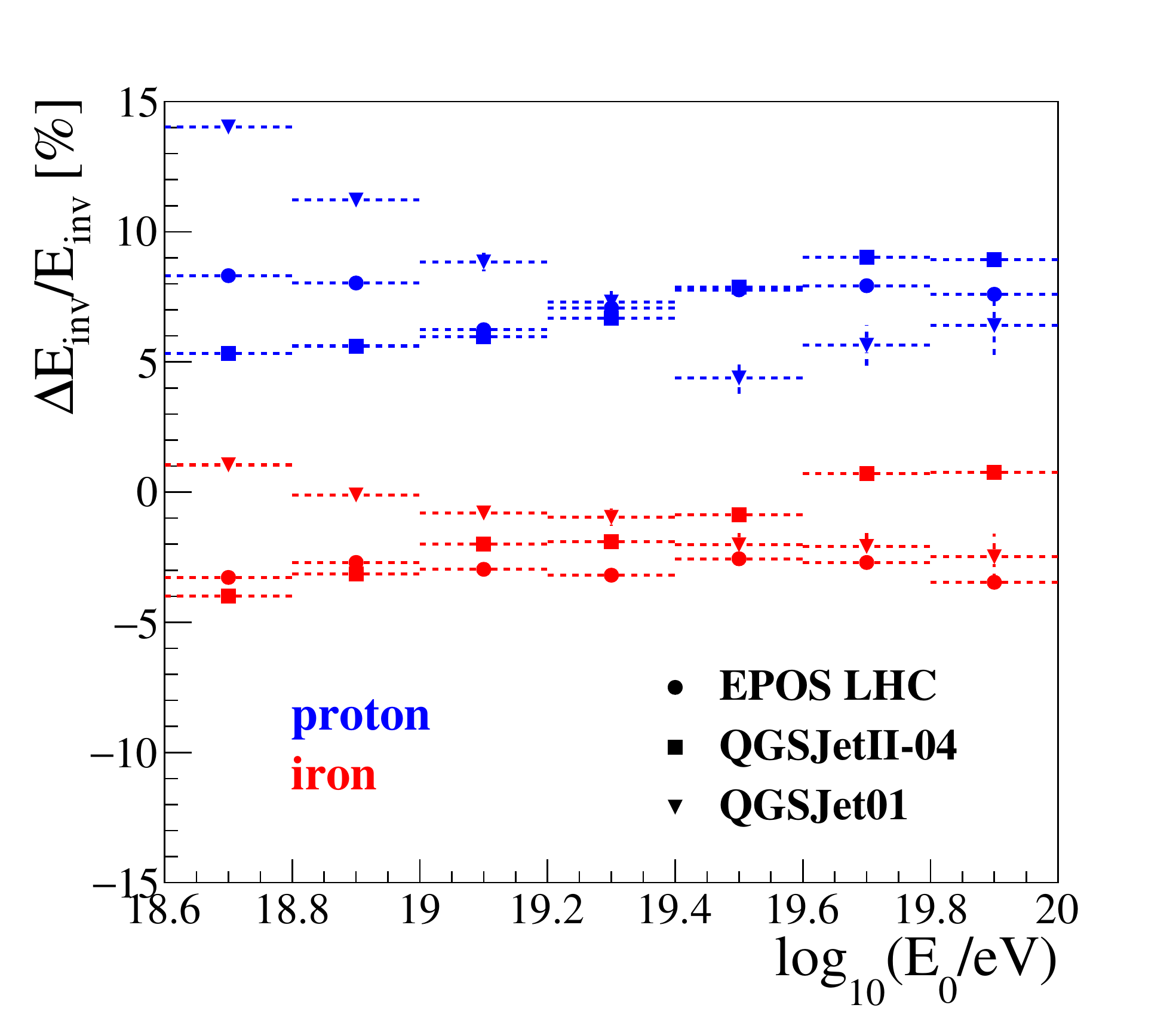}
 }
\caption{
Average value of the relative difference between the reconstructed value of $E_{\rm inv}$ obtained with the
EPOS LHC~\cite{EPOS-LHC}  parametrization for a mixture of 50\% proton and 50\% iron inclined showers. }
\label{fig:EinvBias}
\end{figure}

The performance of the analysis is studied on fully simulated events for which the detector response is simulated
with the same method used to estimate the bias in $N_{19}$~\cite{AugerMuonSize} and $R_{\mu}$ is reconstructed with the same algorithm used for the data. For each simulated event, we compute $E_{\rm inv}$ from $R_{\mu}$ using the estimation for the mixed proton and iron composition, and we compare it with the true value of the invisible energy. The average values of the residuals as a function of the true value of  $E_{\rm inv}$ are shown in Fig.~\ref{fig:EinvBias} for all primaries and hadronic models of table ~\ref{tab:Einv_Rmu}. The residuals are within $\pm 10\%$ which is an indication of the overall systematic uncertainty in the $E_{\rm inv}$ estimation, which is dominated by the model and mass composition dependence of the values of $C$ and $\delta$.

\subsection{$E_{\rm inv}$ from vertical showers}
\label{Sec:Einv_S1000}

As seen in Sec.~\ref{Sec:EinvPhenomenology}, the invisible energy depends on primary energy through a power law relationship
\begin{equation}
E_{\rm inv} =  \epsilon_c^{\pi}  \beta_0  \left(\frac{E_0}{\epsilon_c^{\pi}}\right)^\beta.
\label{eq:Einv_E0_beta0}
\end{equation}
The parameter $\beta_0$, equal to $A^{1-\beta}$ in the Heitler model extended to hadronic cascades~\cite{Matthews} (see Eq.~\eqref{eq:EinvE0A}), has been introduced in order to account for the large variations in the predictions of the number of muons that are obtained using different hadronic interaction models once the shower energy and primary mass are fixed.

In the reconstruction of vertical events, the primary energy is estimated from $S(1000)$ by correcting for the shower attenuation using the constant intensity cut method~\cite{CIC}. To estimate $E_{\rm inv}$ from $S(1000)$, we use the functional form
\begin{equation}
 E_{0}= \gamma_0(\Delta X)\,\left[S(1000)\right]^{\gamma},
\label{eq:E0_CIC}
\end{equation}

\begin{figure*}[]
\centering
 \centerline{
\includegraphics[width=1\columnwidth,height=0.9\columnwidth,clip]{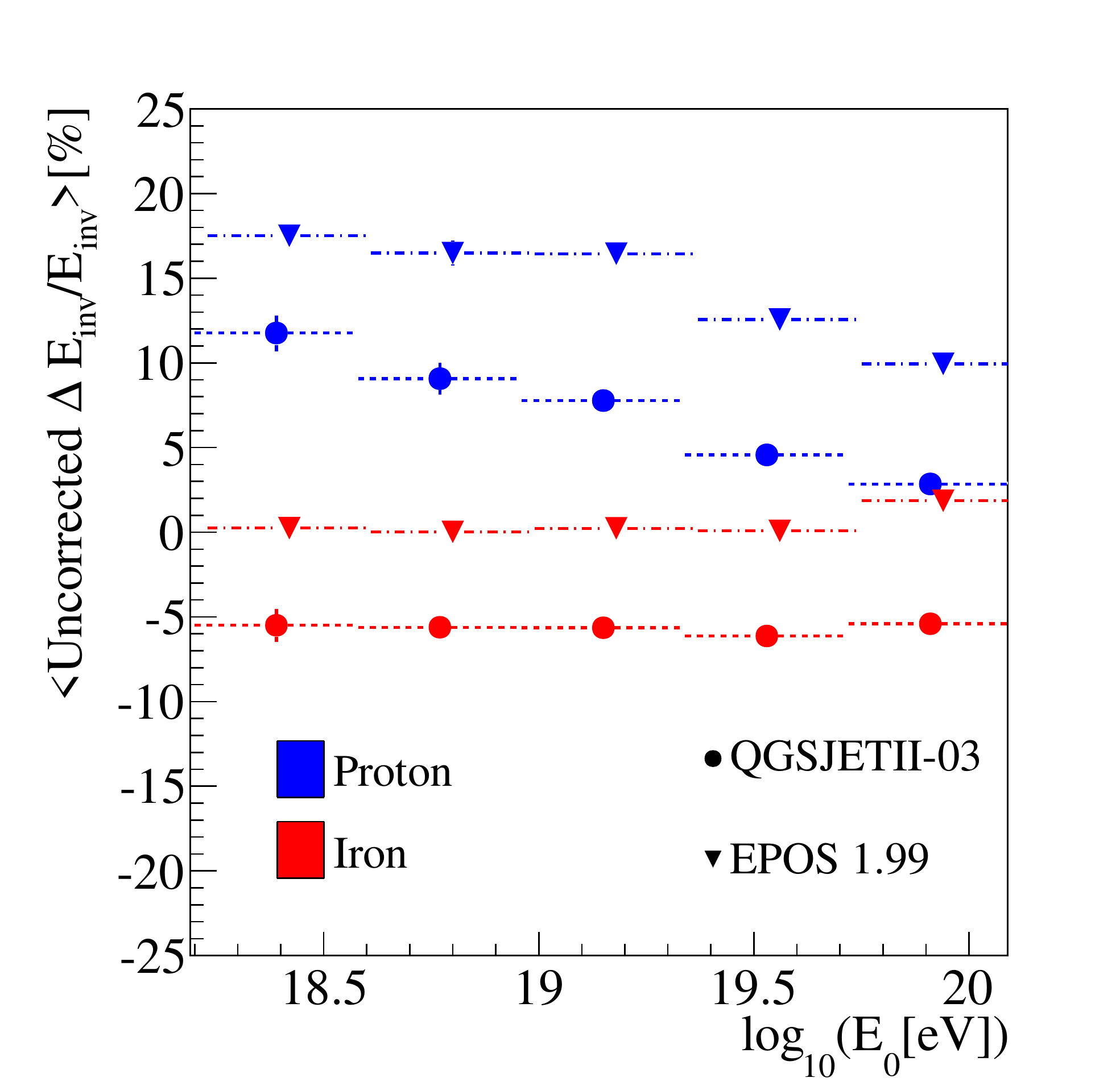}
\includegraphics[width=1\columnwidth,height=0.9\columnwidth,clip]{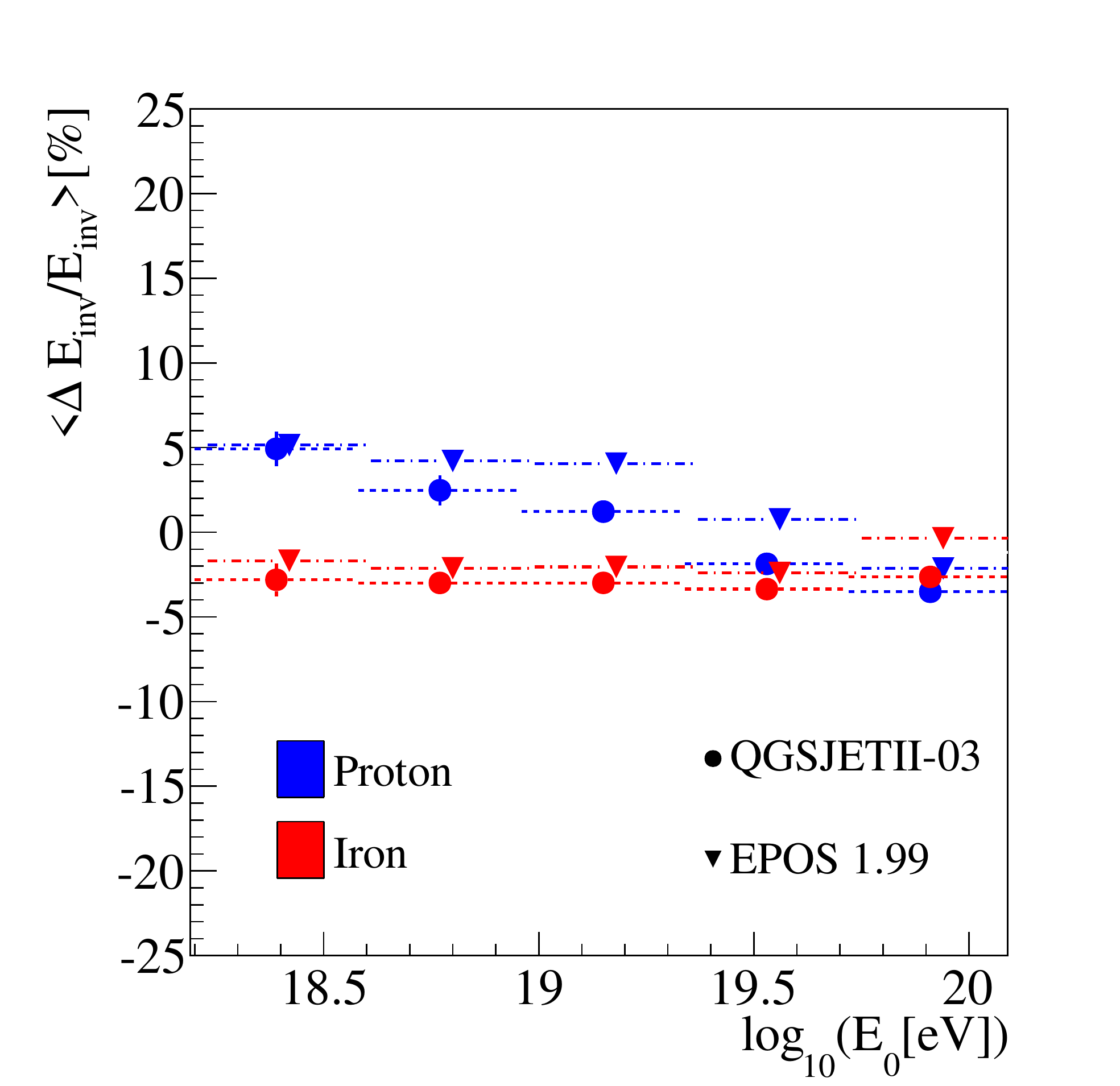}
}
\caption{Average value of the relative difference between the true value of $E_{\rm inv}$ and the value of $E_{\rm inv}$ reconstructed from $S(1000)$ and $X_{\rm max}$ using Eq.~\eqref{eq:Einv_S1000} (left) and after applying the correction due to the difference among the simulations in the predictions of the number of muons and in the attenuation function, using Eq.\eqref{eq:correction} (right). The simulations refer to vertical showers.}
\label{fig:EinvS1000Bias}
\end{figure*}

\noindent
where $\Delta X = 875~{\rm g/cm}^2 /\cos\theta - X_{\rm max}$ is the atmospheric slant depth between ground level and the depth of the shower maximum development ($875~{\rm g/cm^2 }$ is the vertical atmospheric depth of the Auger site, $\theta$ is the zenith angle of the shower), and $\gamma_0(\Delta X)$ is related to the attenuation of $S(1000)$ with $\Delta X$. In contrast with the relationship used to get the SD energies, the dependence of $E_0$ on $\Delta X$ allows us to better take into account the shower-to-shower fluctuations and thus obtain a more precise estimation of $E_{\rm inv}$.

Combining Eq.~\eqref{eq:Einv_E0_beta0} and \eqref{eq:E0_CIC} one obtains
\begin{eqnarray}
E_{\rm inv} & = & \epsilon^{\pi}_{c} ~\beta_0 ~\left(\frac{\gamma_0(\Delta X)\,S(1000)^{\gamma}} {\epsilon^{\pi}_{c}}\right)^{\beta} \\
 & = & A(\Delta X) ~\left[S(1000)\right]^{B},
\label{eq:Einv_S1000}
\end{eqnarray}
where
\begin{eqnarray}
\label{eq:A_DX}
A(\Delta X) & = &  \left( \epsilon^{\pi}_{c} \right)^{1-\beta}   ~\beta_0 ~\left[ \gamma_0(\Delta X) \right]^\beta, \\
B & = & \gamma \beta~.
\label{eq:B}
\end{eqnarray}
The parameter $B$ and those defining the function $A(\Delta X)$ are determined using Monte Carlo simulations. Using the \qgsjet II-03 hadronic interaction model, we find $\beta=0.925$ and $\gamma=1.0594$, so that their product is $B=0.98$. We have verified that different interaction models yield the same value of $B$ to within 2\%. This value will be used from now on, so that with Eq.~\eqref{eq:Einv_S1000} and the measurements of $S(1000)$ and $\Delta X$ one can obtain an event-by-event estimate of $E_{\rm inv}$.

The function $A(\Delta X)$ is calculated using Monte Carlo simulated events in which the WCD is simulated with \textsc{Geant4}~\cite{Geant4} using the Auger \mbox{$\overline{\textrm{Off}}$\hspace{.05em}\protect\raisebox{.4ex}{$\protect\underline{\textrm{line}}$}} software framework~\cite{Offline}. The simulations are done with the \qgsjet II-03~\cite{QGSJet2} hadronic interaction model for a mixed composition of 50\% protons and 50\% iron. Note that $A(\Delta X)$ is parametrized with the fourth-degree polynomial reported in appendix~\ref{Appendix_A}.

The performance of the analysis is tested with Monte Carlo proton and iron events simulated with the hadronic interaction models \qgsjet II-03~\cite{QGSJet2}, and EPOS 1.99~\cite{EPOS} and using the \mbox{$\overline{\textrm{Off}}$\hspace{.05em}\protect\raisebox{.4ex}{$\protect\underline{\textrm{line}}$}} framework to simulate the detector response. Here $~E_{\rm inv}$ is calculated from $S(1000)$ and $\Delta X$ and compared with the true invisible energy. The average values of the residuals as a function of the total shower energy are shown in the left panel of Fig.~\ref{fig:EinvS1000Bias} and are between $-5 \%$ and $20 \%$. The spread in the residuals is mainly due to the difference in the predictions of the number of muons and of the  attenuation function $\gamma_0(\Delta X)$
among the simulations used to parametrise $A(\Delta X)$, and the ones used to simulate the events. We note that the function $\gamma_0(\Delta X)$ includes the conversion factor needed to obtain the shower energy from $S(1000)$ which is strongly model dependent.

A better estimation of $\einv$ can be obtained by taking into account these differences using the following equation
\begin{equation}
 E_{\rm inv}  = A(\Delta X) ~\left[S(1000)\right]^{B} ~\left(   \frac{ \tilde{\gamma}_0(\Delta X)}{    \gamma_0(\Delta X)    }  \right)^\beta  ~\frac{\tilde{\beta}_0}{\beta_0},
\label{eq:correction}
\end{equation}
where the quantities with and without the accent tilde are calculated for the data sample that we are analizing and for the one used to parametrise $A(\Delta X)$, respectively. Here, $\beta$ is fixed to 0.925. The functions $\gamma_0$ are obtained from Eq.~\eqref{eq:E0_CIC} using the shower energy and $S(1000)$. The ratio $\tilde{\beta}_0/\beta_0$ is estimated from the ratio of the number of muons at ground level for the two data sets, information that is available in the CORSIKA events. The residuals in $\einv$ using the improved parametrization of Eq.~\eqref{eq:correction} are shown in the right panel of Fig.~\ref{fig:EinvS1000Bias}. Note that now we can recover the true value of the invisible energy within a few \% for all models and primaries. Note also how we improve the estimation of $\einv$ for \qgsjet II-03, despite the primary mass composition used to parametrise $A(\Delta X)$ being different to that of the simulated events used to test the analysis method. In the next section, we will see how the  parametrization of Eq.~\eqref{eq:correction} is used to estimate the invisible energy of the Auger data.

\subsection{$E_{\rm inv}$ from Auger data and its parametrization as a function of the calorimetric energy}
\label{Sec:Einv_data}

The analysis methods described in Sec.~\ref{Sec:Einv_Rmu} and~\ref{Sec:Einv_S1000} allow us to obtain an event-by-event estimation of $E_{\rm inv}$ for the data collected by the Pierre Auger Observatory. We recall that the vertical events are those with zenith angles $\theta < 60^\circ$, while inclined events have $60^\circ < \theta < 80^\circ$. For both data sets, the analysis is limited to those events sufficiently energetic to ensure a full trigger efficiency. In fact, at lower energies the trigger is biased towards events with a higher number of muons, and thus higher invisible energy and consequently larger systematic uncertainties. The energy thresholds for the full trigger efficiency are $4 \times 10^{18}$ eV for the inclined \cite{HASRec} and $3 \times 10^{18}$ eV for the vertical events \cite{SDExp}.

In order to obtain an invisible energy estimation that can be used for all events detected by the FD, including the ones with energies below the full SD trigger efficiency, the event-by-event estimation of $E_{\rm inv}$ is parametrized  as a function of the calorimetric energy above the full trigger efficiency, with the function being extrapolated to lower energies.

The parametrization is obtained by analizing a sample of hybrid showers selected with the same selection criteria used for the energy calibration of the SD energy estimators~\cite{Auger-EnSc}.

SD events are selected requiring that the WCD with the highest signal is enclosed within a hexagon of six
active stations. This is the basic cut used to calculate the aperture of the SD~\cite{SDExp}, and it rejects events that can be affected by large uncertainties because they fall near the edge of the array or in regions where a station is temporarily not fully operational.

FD events are selected in order to guarantee a precise reconstruction of the longitudinal profile. Good atmospheric conditions are ensured by requiring that the vertical aerosol optical depth is measured within 1 hour of the time of the event and its value at 3 km above ground level is less than 0.1. Moreover, information from the Auger infrared cloud cameras, laser facilities, LIDAR stations and the Geostationary Operational Environmental Satellites (GOES) database are used to discard events detected by telescopes that have clouds in their fields of view~\cite{Auger-mass}. Next, a set of quality selection cuts is applied to obtain a precise reconstruction of the energy deposit $dE/dX$. The total track length  (that is the entire range of depths along which $dE/dX$ is measured)  must be at least $200~{\rm g/cm^2}$. Events are rejected if there is a gap in the energy deposit profile larger than 20\% of the total track length. The  error on the reconstructed calorimetric energy must be less than 20\% and the residual in the Gaisser-Hillas fit, $\left( \chi^2 - n_{\rm dof} \right) / \sqrt{2 n_{\rm dof}}$, must be less than 3. Finally, a set of cuts related to the field of view of the telescopes is applied. The field of view is defined by lower ($X_{\rm l}$) and upper ($X_{\rm u}$) depth boundaries and must be large enough to have equal sensitivity to an appropriately large range of values of the depth of shower maximum $X_{\rm max}$. The cut ensures that the maximum accepted uncertainty in $X_{\rm max}$ is $40~{\rm g/cm^2}$ ($150~{\rm g/cm^2}$) and that the  minimum viewing angle of light in the telescope is $20^\circ$ ($25^\circ$) for the vertical (inclined) showers. Furthermore, the values of $X_{\rm l}$ and $X_{\rm u}$ are to be within certain limits in order to enclose the bulk of the $X_{\rm max}$ distribution. The overall purpose of field-of-view cuts is to select primaries with different masses with the same probability~\cite{Auger-mass}.

A last cut is applied to the FD energies and ensures that the SD trigger efficiency is close to 100\%. Since at this stage of the analysis the invisible energy is not known, the cut is applied to the calorimetric energies, requiring that they are larger than $3.5 \times 10^{18}$ eV for the inclined showers and $2.5 \times 10^{18}$ eV for vertical events. These correspond to the energy thresholds for full trigger efficiency assuming that the invisible energy is about 15\% of the total shower energy \cite{Spectrum-ICRC2017}.

The analysis is performed over the hybrid events collected from 1 January 2004 to 31 December 2015, and the selected data set consists of 310 inclined and 2827 vertical events.

As seen in Sec.~\ref{Sec:Einv_S1000}, an unbiased estimate of $E_{\rm inv}$ in vertical showers requires the two corrections shown in Eq.~\eqref{eq:correction}, one related to the attenuation ($\gamma_0(\Delta X)$) and the other to the muon number ($\beta_0$). The ratio $\tilde{\gamma}_0(\Delta X) / \gamma_0(\Delta X)$ is obtained by doing a fit to the hybrid data (ratio of $S(1000)^\gamma$ to the FD energy in bins of $\Delta X$) to extract $\gamma_0(\Delta X)$. As the statistics are too limited below $\Delta X = 250~{\rm g/cm^2}$ and above $\Delta X = 1000~{\rm g/cm^2}$ for sufficiently small bins of $\Delta X$, and considering that the uncertainty on the FD energy becomes larger when the maximum of the energy deposit is close to ground level, the applicability of the method is presently limited to this range. Thus, a further cut on $\Delta X$ is applied to the vertical events requiring it to be in the above range and, from among the 2827 events, 2389 are selected. The correction factor $\tilde{\beta}_0/\beta_0$ is estimated from the ratio of the average muon number measured in inclined events and the muon number predicted by the model used to calculate the function $A(\Delta X)$ (\qgsjet II-03~\cite{QGSJet2} for a mixed composition of 50\% protons and 50\% iron). Further details on the calculation of the two corrections are reported in appendix~\ref{Appendix_A}. The two corrections are rather large, but they partially compensate for each other. The average value of $\left(\tilde{\gamma}_0/\gamma_0\right)^\beta$ ($\beta = 0.925$) is about 0.73 and almost the same at all energies. It essentially reflects the mismatch between the energy estimation provided by the simulations and that given by the fluorescence measurements. The correction $\tilde{\beta}_0/\beta_0$ is about 1.55 and slightly increases with energy. Despite their large values, the overall correction is only about 1.15.

\begin{figure*}[]
\centering
 \centerline{
\includegraphics[width=1\columnwidth,height=1\columnwidth,keepaspectratio,clip]{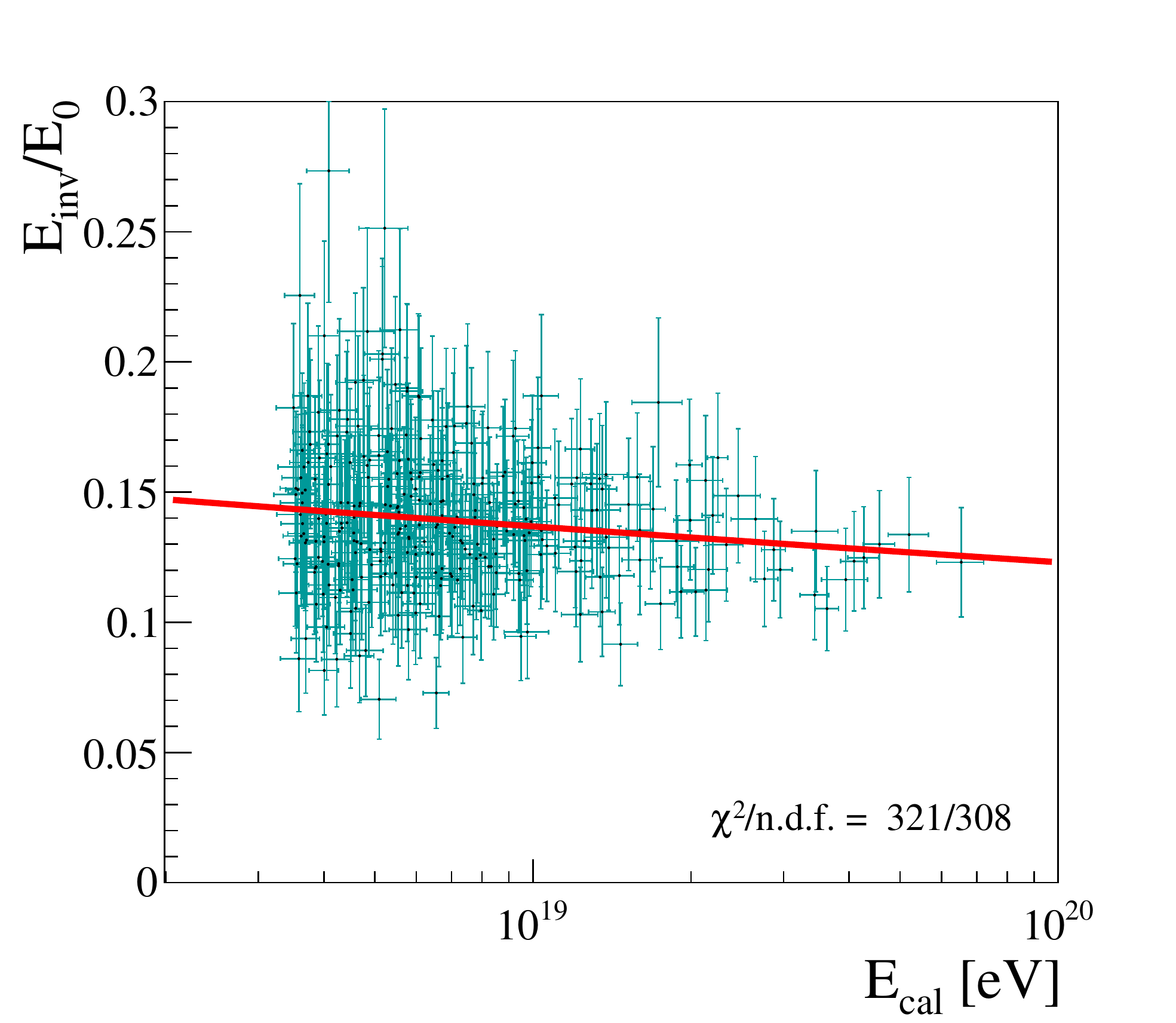}
\includegraphics[width=1\columnwidth,height=1\columnwidth,keepaspectratio,clip]{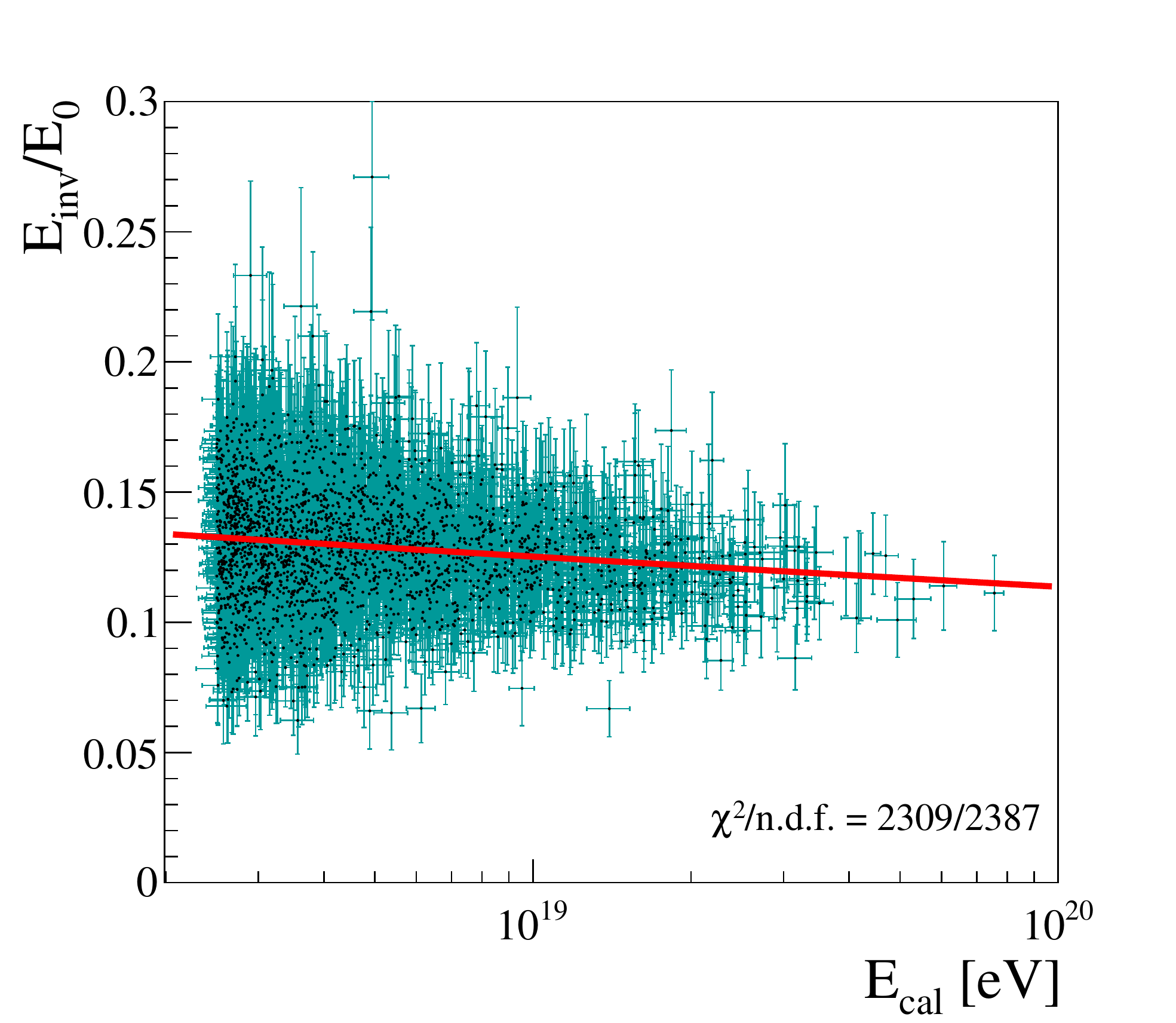}
 }
\caption{Invisible energy fraction from the hybrid data as a function of the calorimetric energy. The estimations obtained with inclined and vertical events are shown in the left and right figures, respectively. Each point represents an individual event and the error bars are calculated propagating the uncertainties in $E_{\rm cal}$ and $R_{\mu}$.
The results of a fit of a power law functional form are shown with red lines.
}
\label{fig:EinvData}
\end{figure*}

The correlation between $E_{\rm inv}$ and $E_{\rm cal}$ is well approximated by a power law relationship
\begin{equation}
E_{\rm inv} = a \left( \frac{E_{\rm cal}}{10^{18} {\rm~ eV} } \right)^b
\label{eq:Einv_Ecal}
\end{equation}
where the parameters $a$ and $b$ are fitted to the data using a maximum-likelihood method~\cite{Spectrum-ICRC2017} that allows us to correctly take into account the cut on energy for the full trigger efficiency, avoiding the bias that would be introduced by a standard least-squares fit.  The calculation of the probability density function includes the event-by-event uncertainties on $E_{\rm cal}$ and $E_{\rm inv}$. A detailed description of the uncertainty on $E_{\rm cal}$ is reported in~\cite{Auger-EnSc}. The uncertainties on $E_{\rm inv}$ for the inclined events are obtained by propagating the event-by-event errors on $R_{\mu}$ as given by the fit of the muon number density~\cite{HASRec}. For the vertical events, the uncertainties arising from $S(1000)$ are calculated as reported in~\cite{Spectrum-ICRC2017} and these dominate over the contribution given by the errors on $X_{\rm max}$. The uncertainties on the invisible energy include a contribution from the shower-to-shower fluctuations with a value inferred from the data, requiring that the reduced $\chi^2$ is approximately equal to 1. This uncertainty amounts to 14\% and 11\% for the inclined and vertical data sets, respectively.

The results of the fit are shown in table~\ref{tab:Einv-Ecal-par} and the data and the fitted functional forms are shown in Fig.~\ref{fig:EinvData}. The data are presented in terms of the ratio $E_{\rm inv} / E_{0}$ (with $E_{0} = E_{\rm cal} + a~(E_{\rm cal}/{\rm EeV})^b $) to be consistent with the previous figures.

\begin{figure*}[]
\centering
 \centerline{
\includegraphics[width=1\columnwidth,height=1\columnwidth,keepaspectratio,clip]{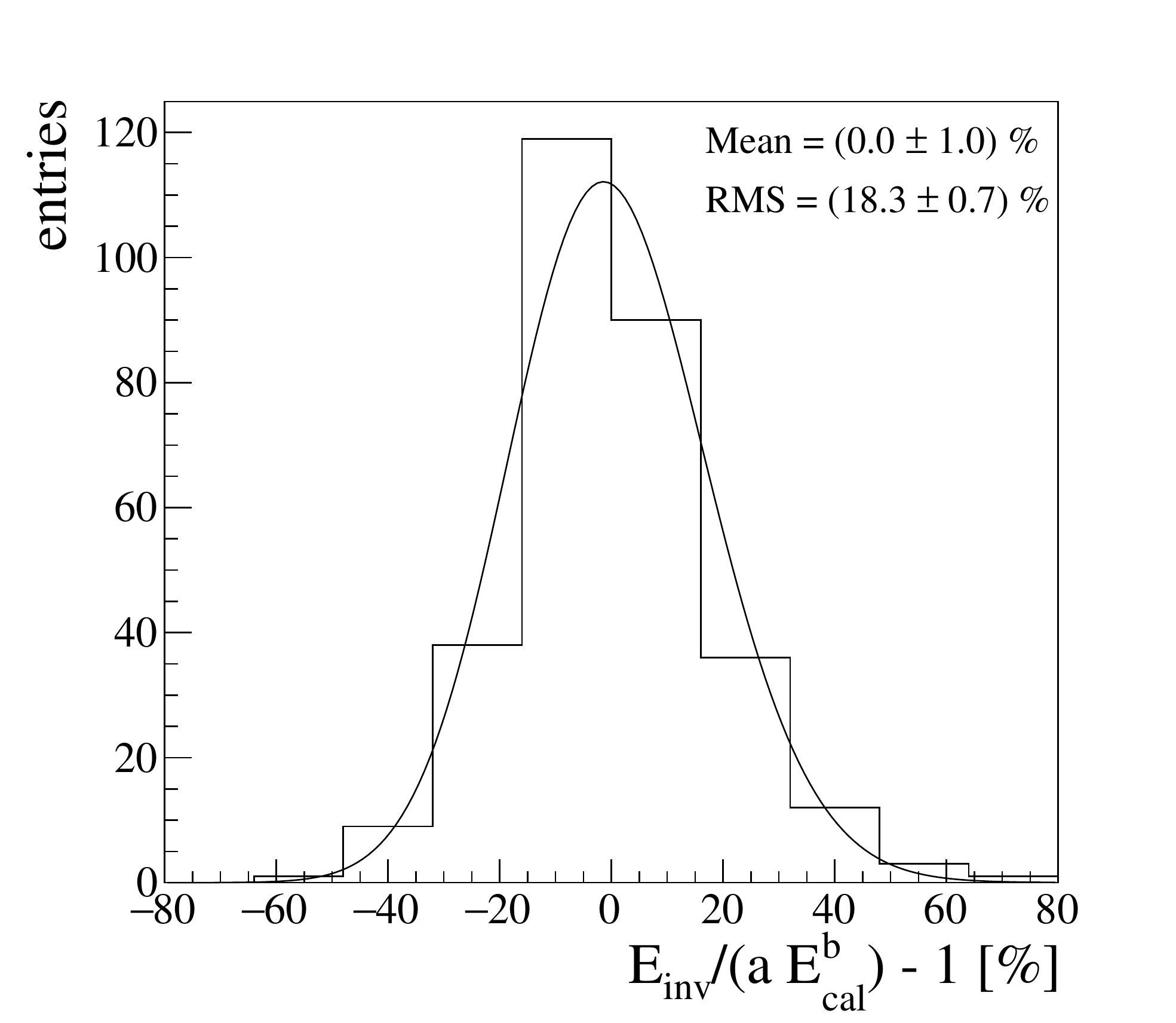}
\includegraphics[width=1\columnwidth,height=1\columnwidth,keepaspectratio,clip]{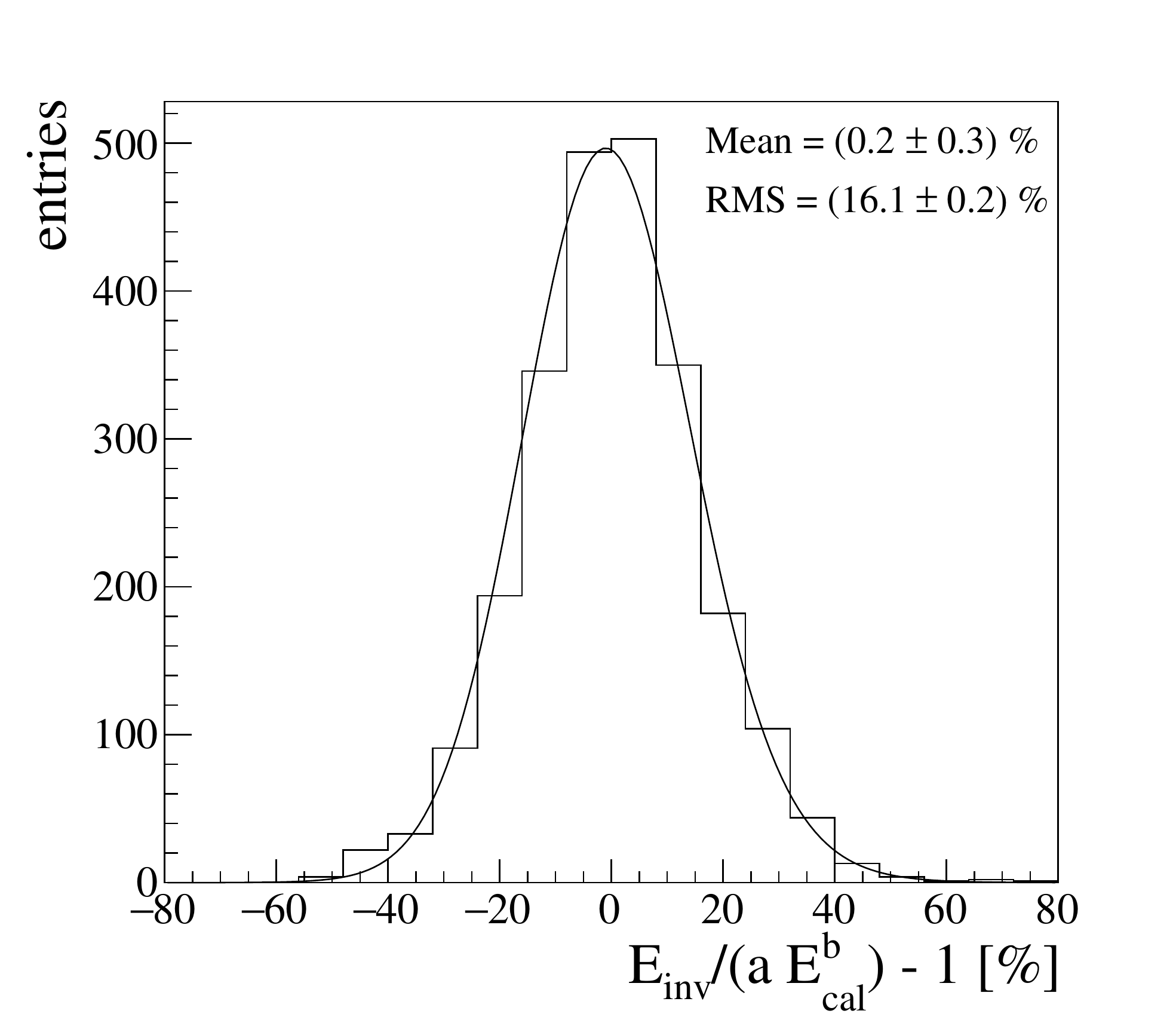}
 }
\caption{Distribution of the residuals between the $E_{\rm inv}$ estimates and the values obtained from $E_{\rm cal}$ using the fitted power law relationship,
 for inclined (left) and vertical (right) events.
The expected distributions for the residuals calculated from the probability model used for the fit are shown with solid lines.
}
\label{fig:EinvRes}
\end{figure*}

Having determined the parameters $a$ and $b$, it is possible to estimate the resolution in $E_{\rm inv}$ from the data. In Fig.~\ref{fig:EinvRes} we show the distribution  of the residuals between $E_{\rm inv}$ and the invisible energy calculated from $E_{\rm cal}$. The width of these distributions is determined by the combined effect of the resolutions on $E_{\rm inv}$ and $E_{\rm cal}$ and then, knowing that the resolution on FD energies is 7.6\%~\cite{Auger-EnSc}, it is possible to estimate the resolution on $E_{\rm inv}$. The fit done using a Gaussian ratio probability density function yields $(16.8 \pm 0.8)\%$ and $(14.4 \pm 0.3)\%$ for inclined and vertical events, respectively, in good agreement with the expected uncertainties.

In the end, in Fig.~\ref{fig:EinvRes} we show that the residual distributions are in good agreement with the expected ones (shown with solid lines) calculated from the probability model used to fit the  $a$ and $b$ parameters, further demonstrating the correctness of the estimation of the uncertainties.

\begin{table}[h]
\caption{Values of the parameters of Eq.~\eqref{eq:Einv_Ecal} fitted to the data. The errors are statistical and $\rho$ is the correlation coefficient. }
\label{tab:Einv-Ecal-par}
\begin{center}
\begin{tabular}{c|c c c c}
{\rm data sample} & $a ~\left[ 10^{18} {\rm~ eV} \right]$ & $b$ & $\rho$ \\ \hline
{\rm inclined}    & $0.179 \pm 0.006$    &   $0.947 \pm 0.017$   & -0.96 \\
{\rm vertical}    & $0.160 \pm 0.002$    &   $0.952 \pm 0.005$   & -0.94 \\
\end{tabular}
\end{center}
\end{table}

\subsection{Systematic uncertainties}

As seen in Sec.~\ref{Sec:Einv_Rmu}, the analysis method to estimate the invisible energy in inclined events allows us to recover the true value of $E_{\rm inv}$ within 10\% for several hadronic interaction models and primary masses (see Fig.~\ref{fig:EinvBias}). The deviations arise from the slight model dependence of the parameters used to get $E_{\rm inv}$ from $R_\mu$ and from the uncertainty in the $R_\mu$ reconstruction. An additional uncertainty related  to the influence of the inclined events close to $60^\circ$, for which the electromagnetic correction is not negligible~\cite{HASRec}, is evaluated using data. Excluding events with zenith angles below $65^\circ$, the fitted parametrization of $E_{\rm inv}$ as a function of $E_{\rm cal}$ changes by less than $2\%$.  Finally, the statistical errors on the $a$ and $b$ parameters cause an uncertainty in $E_{\rm inv}$ below $5\%$.  
The first two uncertainties are both related to the assumption on the primary mass composition and are expected to be partially anticorrelated
(for heavier primaries, we have larger values of $C$ in Eq.~\ref{eq:EinvN19} and a smaller electromagnetic correction that causes a decrease of $C$). Thus, a conservative
estimate of a total uncertainty of about 12\% is obtained neglecting such correlation and adding all contributions in quadrature.

Concerning the vertical events, we have seen in Sec.~\ref{Sec:Einv_S1000} that the analysis method allows
us to recover the correct $E_{\rm inv}$ well within 5\%, that we conservatively consider as a systematic uncertainty. Other contributions to the uncertainty arise from the statistical error on the $a$ and $b$ parameters ($<2 \%$), from the systematic uncertainty in the $X_{\rm max}$ scale ($1\%$) and from the uncertainty in the $\beta_0$ correction (the statistical error on the parametrization of $N_{19}$ as a function of energy and systematics related to the electromagnetic correction, with an overall contribution of less than $4\%$). Another uncertainty comes from a possible correction for the exponent $B$ of the power law relationship between $E_{\rm inv}$ and $S(1000)$ (see Eq.~\eqref{eq:Einv_S1000}). Changing $B$ by $2\%$, we estimate an uncertainty in $E_{\rm inv}$ that increases with energy from $5\%$ to $11\%$.
A total uncertainty of $11\% - 16\%$ (larger at higher energies) is obtained assuming that the errors on  $B$ and the 5\% addressed in Sec.~\ref{Sec:Einv_S1000} are fully correlated (both depend on the model and primary mass used to estimate the parameters) and noting that the other uncertainties are expected to be largely uncorrelated.

\section{Discussion of the results and extrapolation to low energies}
\label{Sec:Einv_discussion}

For a quantitative comparison of the two data-driven estimations of $\einv$ presented in the previous section one has to take into account that $\einv$ has a zenith angle dependence, being larger for showers at larger zenith angles. The zenith angle dependence of $\einv$ has been studied simulating proton and iron showers at $10^{18.5}$ eV using the AIRES~\cite{AIRES} code with the \qgsjet II-03~\cite{QGSJet2} hadronic interaction model and parametrized with an analytical function that is reported in appendix~\ref{Appendix_B}. We find that the average values of $\einv$ for the two data sets differ by 5\%: $E_{\rm inv}(41^\circ)/E_{\rm inv}(66^\circ) \simeq 0.95$ where $66^\circ$ and $41^\circ$ are the average zenith angles of the inclined  and vertical data, respectively. Then, for a correct comparison, one of the two estimates has been corrected by this ratio.  Since the majority of the events have zenith angles below $60^\circ$, in the following we correct the $E_{\rm inv}$ parametrization obtained from the inclined data set,  multiplying the corresponding parametrization as a function of $E_{\rm cal}$ by $0.95$.

\begin{figure*}[]
\centering
 \centerline{
   \includegraphics[width=1\columnwidth,height=1\columnwidth,keepaspectratio,clip] {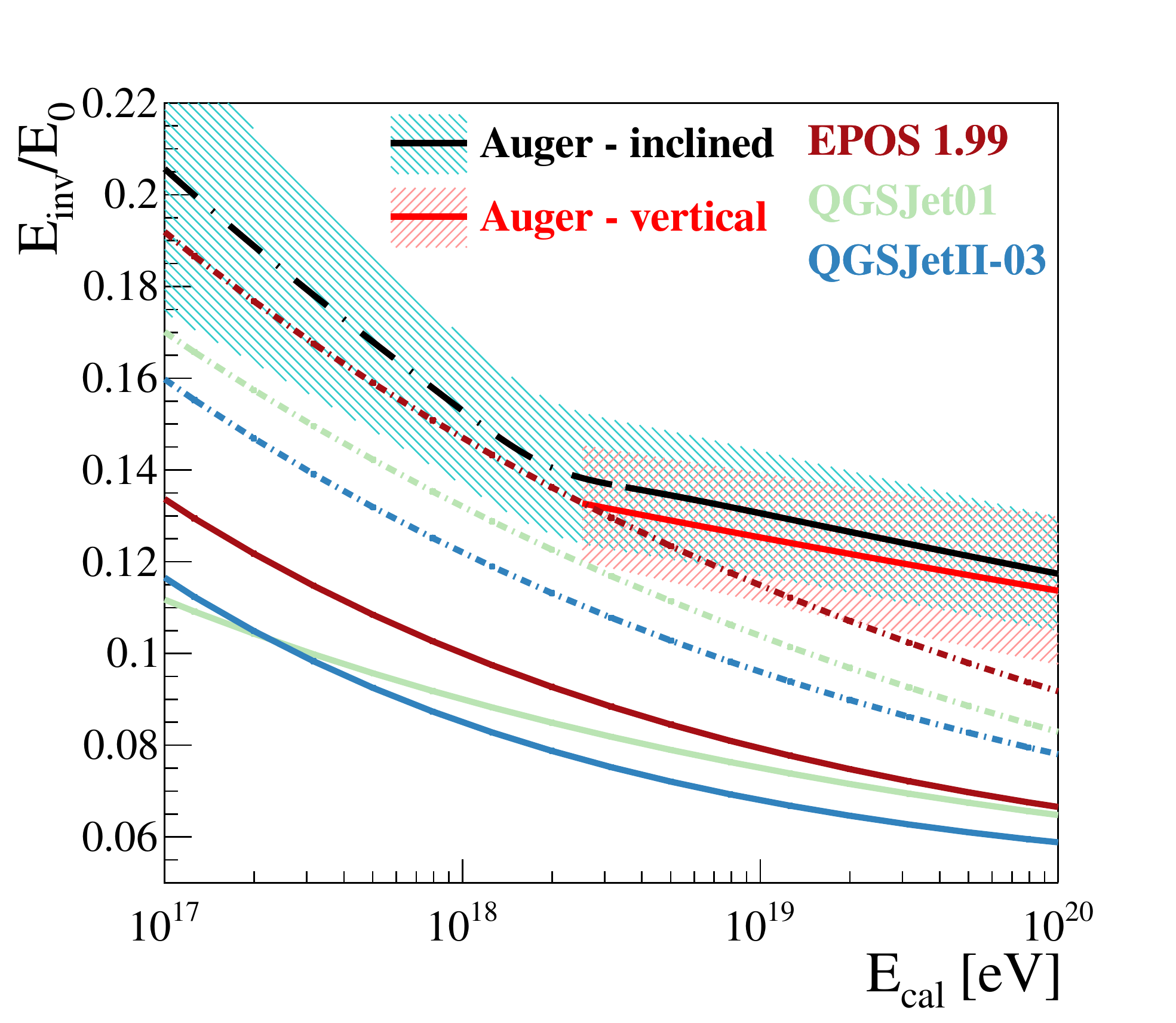}
   \includegraphics[width=1\columnwidth,height=1\columnwidth,keepaspectratio,clip] {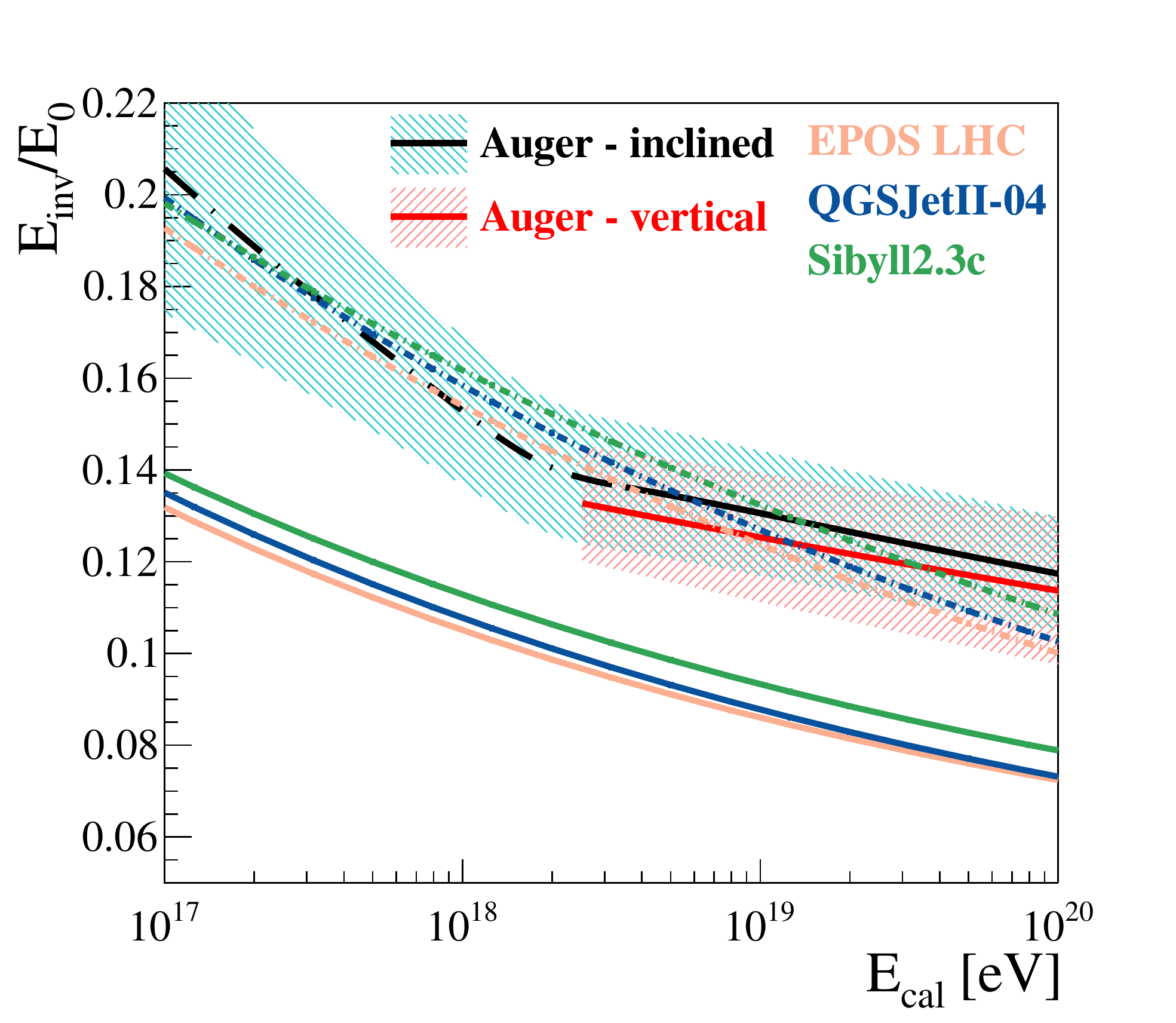}
 }
\caption{Invisible energy obtained for inclined and vertical events compared with the predictions given by Monte Carlo simulations. The estimate for inclined events is extrapolated to low energies, and the systematic uncertainty of both estimations are shown with the shaded bands. The simulations are performed  for proton (solid lines) and iron (dashed lines) primaries and different hadronic interaction models. Left panel: EPOS 1.99~\cite{EPOS}, \qgsjet 01~\cite{QGSJet1} and \qgsjet II-03~\cite{QGSJet2}. Right panel:   EPOS LHC~\cite{EPOS-LHC}, \qgsjet II-04~\cite{QGSJet2-04} and Sibyll2.3c~\cite{Sibyll2.3c}.}
\label{fig:Einv-ecal1}
\end{figure*}

The two data-driven $\einv$ estimations are compared in Fig.~\ref{fig:Einv-ecal1}. They are in excellent agreement, well within the systematic uncertainties that are shown with shaded bands. It is worth noting that the two estimates are partially correlated since they both use the measurement of the muon number. However, they are affected by different systematics and in particular, those arising from the model dependence of the parameters used to get $E_{\rm inv}$ from the shower observables are not expected to be significantly correlated. In fact, in inclined events, the largest model dependence arises from $\epsilon^\pi_c$, while in vertical events, the uncertainty in this parameter only marginally affects the invisible energy since $E_{\rm inv} \propto (\epsilon^\pi_c)^{1-\beta}$ with $\beta \approx 0.9$.

In Fig.~\ref{fig:Einv-ecal1}, we also show the theoretical predictions for the different hadronic interaction models and primary masses addressed in Sec.~\ref{Sec:EinvPhenomenology}. Note that, in comparison to the predictions of the pre-LHC models (left panel), our estimations are in better agreement with the ones of the post-LHC models (right panel). However, they have still  large values, even larger than the predictions for iron primaries, in contradiction with the mean mass composition obtained using the $X_{\rm max}$ measurement~\cite{Auger-mass}.This is a consequence of the muon number deficit in models~\cite{AugerMuonSize}, given that the models fail to describe the properties of shower development related to muons and therefore to $E_{\rm inv}$.

The estimations of $\einv$ have been obtained for energies above the threshold for the full trigger efficiency of the SD. They can be extrapolated to lower energies taking into account the change in the elongation rate (i.e. how the mean mass composition evolves with energy) measured by Auger at $E_{\rm cal}^A \simeq 2 \times 10^{18} \rm{eV}$ \cite{Auger-mass,Auger-mass-comp-ICRC17}. The function is obtained by extrapolating the parametrization obtained from data down to $E_{\rm cal}^A$ and, below this energy, using a model inspired function that matches the parametrization at $E_{\rm cal}^A$.

For the latter, we use the function of Eq.~\eqref{eq:EinvE0A} in which the mean composition as a function of energy is taken from the Auger FD measurements~\cite{Auger-mass-comp-ICRC17} together with a value of $\beta = 0.9$ that reproduces the simulations at lower energies. Then the extrapolated function is approximated by a simple power law function of $E_{\rm cal}$ with the exponent $b_{\rm extr} = 0.846$. The value of $b_{\rm extr}$ reflects how the mass composition evolves with energy, and the particular value chosen for $\beta$.

The extrapolation of $\einv$ obtained from the inclined events is shown with the black dashed line in Fig.~\ref{fig:Einv-ecal1}. A smooth  transition between the power law valid below $E_{\rm cal}^A$ and the one valid at higher energies (shown with the solid black line) is obtained with a hyperbolic tangent function. Further details on the functional shape that parametrises $\einv$ are reported in appendix~\ref{Appendix_B}.

The uncertainty in the extrapolation (shown with the shaded band) is obtained by adding in quadrature the $12\%$ error on the estimate at higher energies and a contribution ($15\%$ at  $10^{17}$ eV that progressively reduces to 0 as we approach $E_{\rm cal}^A$) obtained by changing $\beta$ over a wide range (0.87-0.93).

From Fig.~\ref{fig:Einv-ecal1}, one can see how the rate with which $\einv$ evolves with energy is different above and below $E_{\rm cal}^A$. This is a consequence of the change in the elongation rate. Moreover, the constraint to match our estimation of $E_{\rm inv}$ at $E_{\rm cal}^A$ helps us obtain a more realistic estimation of the extrapolation. However, it is worth noting that this is still an extrapolation that uses measurements of the number of muons at high energy that show an excess with respect to the predictions of the hadronic interaction models. The muon number excess could be different at low energies, and this could introduce additional biases in $E_{\rm inv}$. A more accurate estimation of $\einv$ will be obtained in the near future using the data collected by the AMIGA muon detectors~\cite{AMIGA} installed at the Observatory and using the 750m-spacing sub-array of WCDs~\cite{Auger}.

\section{Conclusions}

We have presented a data-driven estimation of the invisible energy of cosmic ray showers detected by the Pierre Auger Observatory. We have developed two analysis methods for the SD events inclined at zenith angles $60^\circ < \theta < 80^\circ$ and for hybrid showers with $\theta < 60^\circ$. The invisible energy has been parametrized as a function of the calorimetric energy and extrapolated to energies below the full trigger efficiency of the SD.

The two estimations agree at a level well within the systematic uncertainties that are estimated to be of the order of $10\%-15\%$, and give values of $E_{\rm inv}$ considerably higher than the predictions given by Monte Carlo simulations. This is a consequence of the muon number deficit in models~\cite{AugerMuonSize}, a deficit due to the failure of the hadronic interaction models to describe the properties of shower development related to muons. Moreover, the  estimations are consistent with the evolution of the mass composition with energy as measured by Auger~\cite{Auger-mass,Auger-mass-comp-ICRC17}. This is due to the sensitivity of the muon number to the primary mass and, at lower energy, due to the use of the mean mass composition to find the functional form that describes $\einv$ as a function of $\ecal$.

While the two estimations are affected by comparable systematic uncertainties, the one obtained using the inclined events is intrinsically better. In fact, for these showers, we measure the total number of muons arriving at ground level which makes the analysis of $\einv$ rather straightforward, more direct and simpler than the analysis used for vertical events.

\begin{figure}[htb]
\centering
 \centerline{
\includegraphics[width=1\columnwidth,height=1\columnwidth,keepaspectratio,clip]{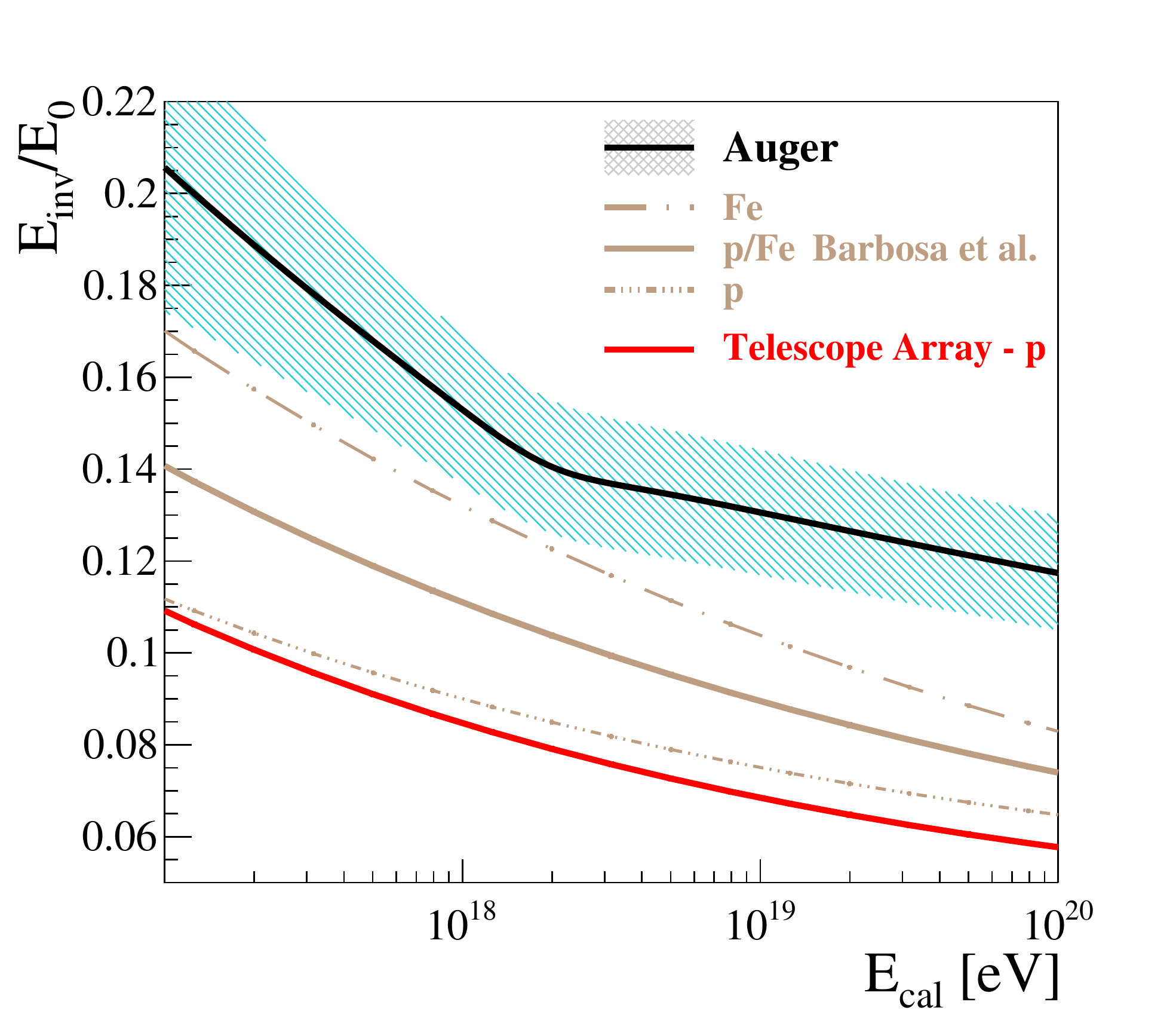}
 }
\caption{Auger data-driven estimation of the invisible energy compared with the parametrizations for protons, iron and mixed composition reported in~\cite{Barbosa} and the one in use  by Telescope Array~\cite{TA-Einv}.}
\label{fig:Einv-ecal2}
\end{figure}

A preliminary data-driven estimation of $\einv$ has already been in use by Auger for several years~\cite{UHECR2016Einv,ICRC2013Einv}. Before 2013, we used a parametrization fully based on simulations assuming  a mixed composition of proton and iron primaries~\cite{Barbosa} that is shown in Fig.~\ref{fig:Einv-ecal2}  and compared with the $\einv$ estimate obtained in this paper from the analysis of inclined events and extrapolated to low energies. In the same figure, we also show the parametrizations obtained in~\cite{Barbosa} for proton and iron primaries and that in use by the Telescope Array Collaboration that assumes a proton composition~\cite{TA-Einv}.  The systematic uncertainty in our estimation is depicted with a shaded band.  From the figure one can evaluate the impact of the data-driven estimation of $E_{\rm inv}$ on the energy scale of the Observatory. Using the two simulations, we  would introduce a bias in the energy scale of $- 4\%$ and $-6 \%$, which is significant considering that the systematic uncertainty in the energy scale introduced by the $E_{\rm inv}$ estimate presented in this paper is about $1.5\%$.

A precise determination of the energy scale is particularly important for the measurement of the energy spectrum of the cosmic rays.
The spectrum falls off with energy approximately as a power-law function ($J \propto E^{-\gamma}$) with a power index $\gamma$ of about 3, and,  at the highest
energies it manifests a flattening at about $5 \times 10^{18}$ eV (feature known as {\it ankle}) and an abrupt suppression above $10^{19}$ eV~\cite{Spectrum-ICRC2017}.
$J$ is measured counting the number of events in bins of energy and it is very sensitive to the energy scale because the systematic uncertainties affecting the
shower energy ($\Delta E/E$) are amplified by the spectral index through the factor $(1-\Delta E/E)^{-\gamma+1}$.
Then one can infer the impact of the data-driven estimation of $E_{\rm inv}$ in the spectrum. Using the two simulations of Fig.~\ref{fig:Einv-ecal2} we could underestimate by about $- 4\%$ and $-6 \%$ the energies at which we observe the spectral features and introduce a negative bias
in the flux $J$ even larger than 10\%.

The invisible energy parametrization as a function of $\ecal$ obtained using the Auger data can also be used in other experiments employing the fluorescence technique. Note that in this case, the uncertainty in $\einv$ remains the one determined in this paper only if the relative calibration factor between the energy scales of the second experiment and Auger is known and taken into account in the calculation of $\einv$. Otherwise, a correct estimation of the uncertainty in $\einv$ needs to take into account the uncertainties in the energy scales of both experiments.


\section*{Acknowledgments}

\begin{sloppypar}
The successful installation, commissioning, and operation of the Pierre
Auger Observatory would not have been possible without the strong
commitment and effort from the technical and administrative staff in
Malarg\"ue. We are very grateful to the following agencies and
organizations for financial support:
\end{sloppypar}

\begin{sloppypar}
Argentina -- Comisi\'on Nacional de Energ\'\i{}a At\'omica; Agencia Nacional de
Promoci\'on Cient\'\i{}fica y Tecnol\'ogica (ANPCyT); Consejo Nacional de
Investigaciones Cient\'\i{}ficas y T\'ecnicas (CONICET); Gobierno de la
Provincia de Mendoza; Municipalidad de Malarg\"ue; NDM Holdings and Valle
Las Le\~nas, in gratitude for their continuing cooperation over land
access; Australia -- the Australian Research Council; Brazil -- Conselho
Nacional de Desenvolvimento Cient\'\i{}fico e Tecnol\'ogico (CNPq);
Financiadora de Estudos e Projetos (FINEP); Funda\c{c}\~ao de Amparo \`a
Pesquisa do Estado de Rio de Janeiro (FAPERJ); S\~ao Paulo Research
Foundation (FAPESP) Grants No.~2010/07359-6 and No.~1999/05404-3;
Minist\'erio da Ci\^encia, Tecnologia, Inova\c{c}\~oes e Comunica\c{c}\~oes (MCTIC);
Czech Republic -- Grant No.~MSMT CR LTT18004, LO1305, LM2015038 and
CZ.02.1.01/0.0/0.0/16\_013/0001402; France -- Centre de Calcul
IN2P3/CNRS; Centre National de la Recherche Scientifique (CNRS); Conseil
R\'egional Ile-de-France; D\'epartement Physique Nucl\'eaire et Corpusculaire
(PNC-IN2P3/CNRS); D\'epartement Sciences de l'Univers (SDU-INSU/CNRS);
Institut Lagrange de Paris (ILP) Grant No.~LABEX ANR-10-LABX-63 within
the Investissements d'Avenir Programme Grant No.~ANR-11-IDEX-0004-02;
Germany -- Bundesministerium f\"ur Bildung und Forschung (BMBF); Deutsche
Forschungsgemeinschaft (DFG); Finanzministerium Baden-W\"urttemberg;
Helmholtz Alliance for Astroparticle Physics (HAP);
Helmholtz-Gemeinschaft Deutscher Forschungszentren (HGF); Ministerium
f\"ur Innovation, Wissenschaft und Forschung des Landes
Nordrhein-Westfalen; Ministerium f\"ur Wissenschaft, Forschung und Kunst
des Landes Baden-W\"urttemberg; Italy -- Istituto Nazionale di Fisica
Nucleare (INFN); Istituto Nazionale di Astrofisica (INAF); Ministero
dell'Istruzione, dell'Universit\'a e della Ricerca (MIUR); CETEMPS Center
of Excellence; Ministero degli Affari Esteri (MAE); M\'exico -- Consejo
Nacional de Ciencia y Tecnolog\'\i{}a (CONACYT) No.~167733; Universidad
Nacional Aut\'onoma de M\'exico (UNAM); PAPIIT DGAPA-UNAM; The Netherlands
-- Ministry of Education, Culture and Science; Netherlands Organisation
for Scientific Research (NWO); Dutch national e-infrastructure with the
support of SURF Cooperative; Poland -- National Centre for Research and
Development, Grant No.~ERA-NET-ASPERA/02/11; National Science Centre,
Grants No.~2013/08/M/ST9/00322, No.~2016/23/B/ST9/01635 and No.~HARMONIA
5--2013/10/M/ST9/00062, UMO-2016/22/M/ST9/00198; Portugal -- Portuguese
national funds and FEDER funds within Programa Operacional Factores de
Competitividade through Funda\c{c}\~ao para a Ci\^encia e a Tecnologia
(COMPETE); Romania -- Romanian Ministry of Research and Innovation
CNCS/CCCDI-UESFISCDI, projects
PN-III-P1-1.2-PCCDI-2017-0839/19PCCDI/2018, PN-III-P2-2.1-PED-2016-1922,
PN-III-P2-2.1-PED-2016-1659 and PN18090102 within PNCDI III; Slovenia --
Slovenian Research Agency; Spain -- Comunidad de Madrid; Fondo Europeo
de Desarrollo Regional (FEDER) funds; Ministerio de Econom\'\i{}a y
Competitividad; Xunta de Galicia; European Community 7th Framework
Program Grant No.~FP7-PEOPLE-2012-IEF-328826; USA -- Department of
Energy, Contracts No.~DE-AC02-07CH11359, No.~DE-FR02-04ER41300,
No.~DE-FG02-99ER41107 and No.~DE-SC0011689; National Science Foundation,
Grant No.~0450696; The Grainger Foundation; Marie Curie-IRSES/EPLANET;
European Particle Physics Latin American Network; European Union 7th
Framework Program, Grant No.~PIRSES-2009-GA-246806; and UNESCO.
\end{sloppypar}

\begin{appendices}

\section{Parameters relevant for the $\einv$ estimation in vertical showers}
\label{Appendix_A}

Both $A(\Delta X)$ ($= E_{\rm inv}/S(1000)^B$, see Eq.~\eqref{eq:Einv_S1000}) and $\gamma(\Delta X)$ ($= E_o/S(1000)^\gamma$, see Eq.~\eqref{eq:E0_CIC}), are parametrized in units of GeV with the function $10^{f(\Delta X)}$ where the exponent is a fourth-degree polynomial in $\Delta X = (875/\cos\theta - X_{\rm max})~{\rm [g/cm^2]}$,
\begin{equation}
f(\Delta X) = p_0 + p_1~\frac{\Delta X}{1000} + p_2 \left( \frac{\Delta X}{1000} \right)^2  + p_3 \left( \frac{\Delta X}{1000} \right)^3
\label{eq:A_gamma_DX_fun}
\end{equation}
and where the values of the parameters are shown in table~\ref{tab:A(X)}.

\begin{table}[h]
\caption{Values of the parameters that define the functions $A(\Delta X)$ and $\gamma_0(\Delta X)$ (see Eq.~\eqref{eq:A_gamma_DX_fun}).}
\label{tab:A(X)}
\begin{center}
\begin{tabular}{cc|c c c c}
                                          &                                            & $p_0$ & $p_1$ & $p_2$  & $p_3$ \\ \hline
  $A(\Delta X)$                           & {\small \qgsjet II-03~\cite{QGSJet2}}\\
                                          & - 50\% p 50\% Fe                           & 7.396 & -0.696 & 2.310 & -1.224 \\ \hline
 \multirow{ 3}{*}{$\gamma_0(\Delta X)$ }  & {\small \qgsjet II-03~\cite{QGSJet2} - p}  & 8.363  & -0.651 & 2.486 & -1.353 \\
                                          & {\small \qgsjet II-03~\cite{QGSJet2} - Fe} & 8.215  & -0.293 & 1.829  & -1.034 \\
                                          & hybrid data                                & 8.306  & -0.984 & 2.644  & -1.454  \\
\end{tabular}
\end{center}
\end{table}

\begin{figure*}[]
\centerline{
\includegraphics[width=0.98\columnwidth,height=0.98\columnwidth,keepaspectratio,clip]{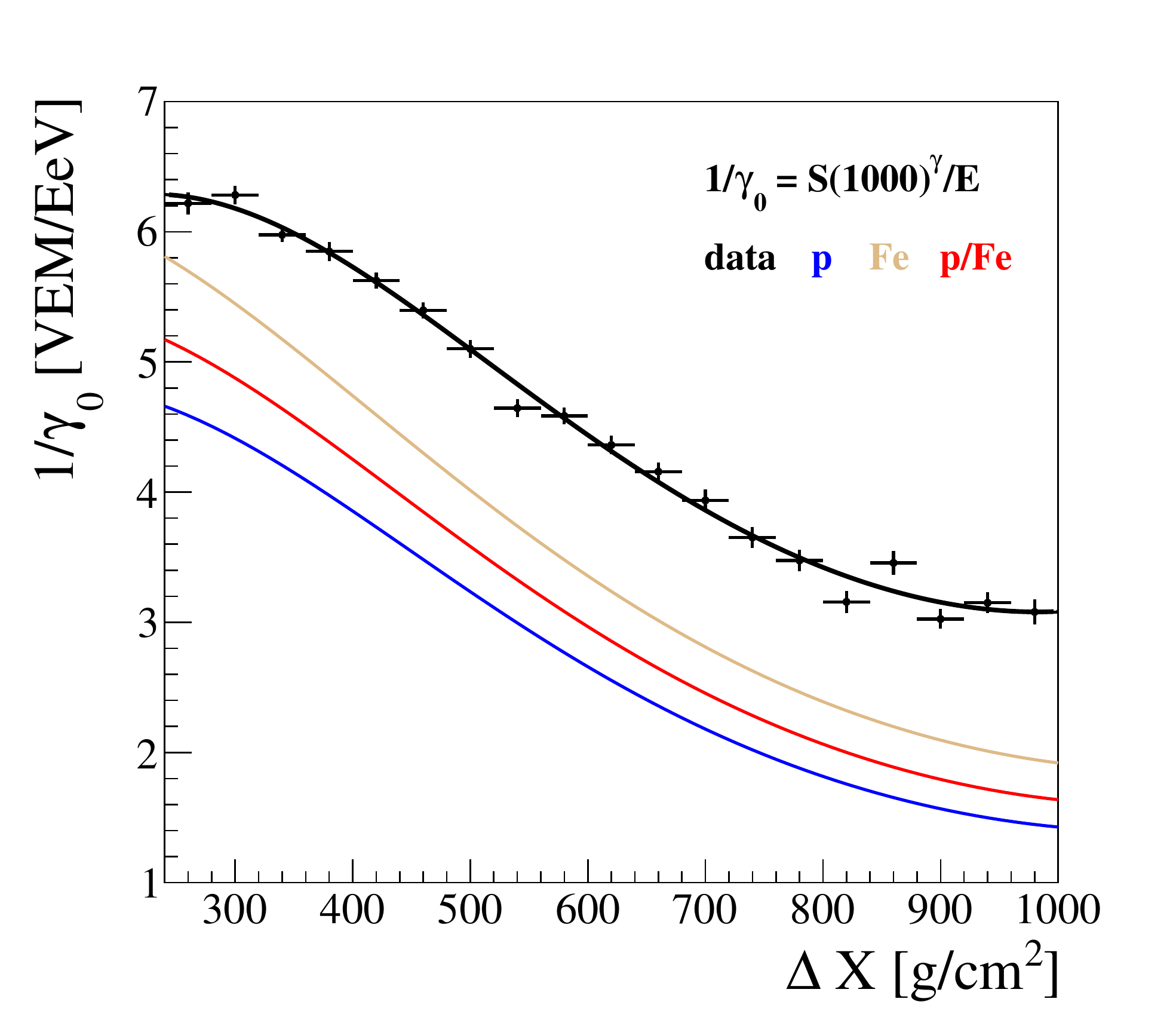}
\includegraphics[width=0.98\columnwidth,height=0.98\columnwidth,keepaspectratio,clip]{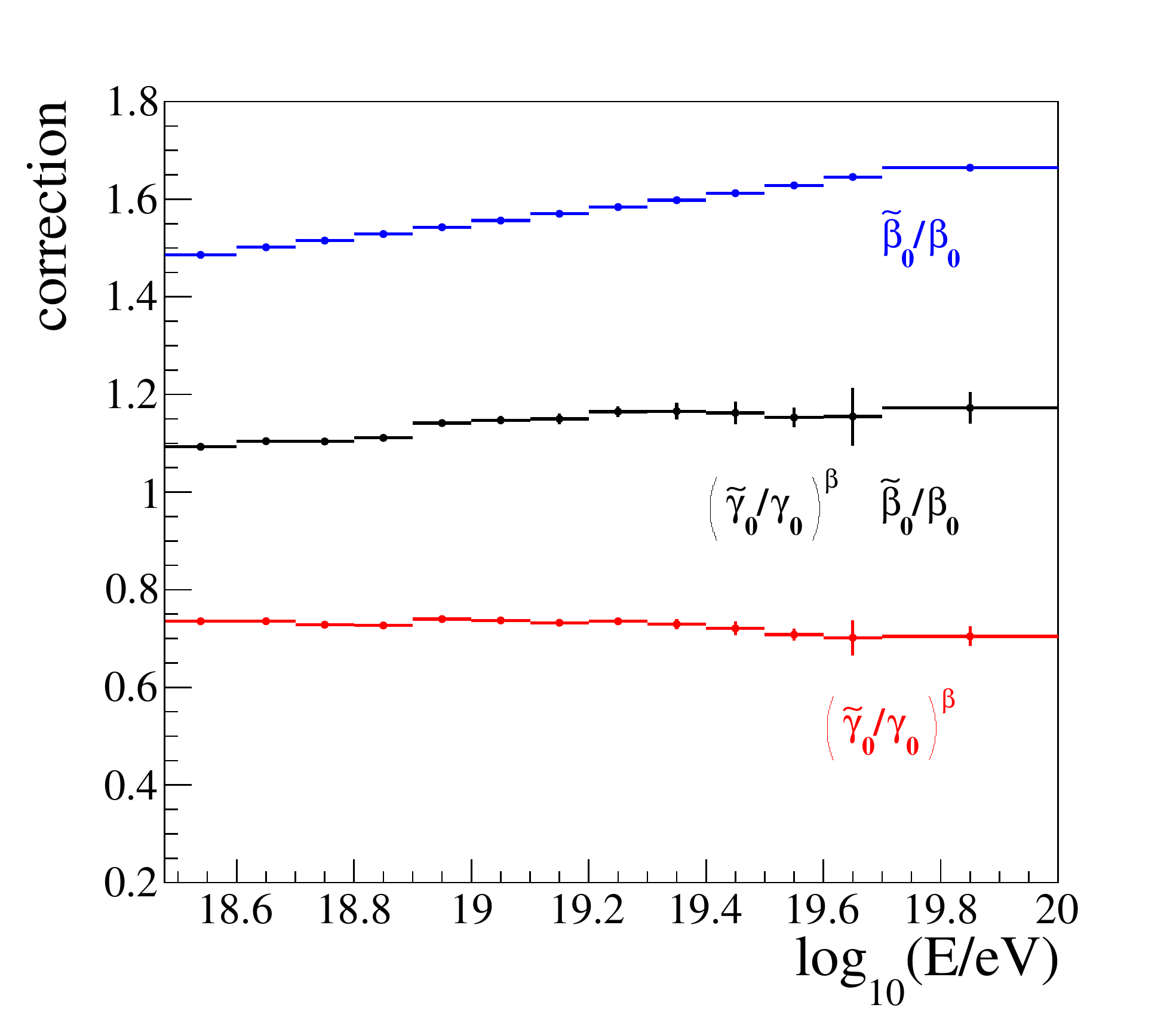}
 }
\caption{Left: the inverse of $\gamma_0(\Delta X)$ as predicted by the \qgsjet II-03~\cite{QGSJet2} simulations and as parametrized using the hybrid data. Right: mean values of the correction factors needed to estimate $E_{\rm inv}$ from data calculated in bins of shower energy.}
\label{fig:EinvCorr}
\end{figure*}

The function $\gamma_0(\Delta X)$ for a mixed composition of $50\%$ p - $50\%$ Fe is calculated taking the average of the functions $\gamma_0$ for proton and iron primaries.  All the parametrizations, together with the average values of $S(1000)^\gamma/E$ in bins of $\Delta X$ obtained with the hybrid data are shown in the left panel of Fig.~\ref{fig:EinvCorr}. Note that $\gamma_0(\Delta X)$ for data is calculated by assuming $\gamma = 1.0594$, the value that reproduces the \qgsjet II-03 ~simulations (see Sec.~\ref{Sec:Einv_S1000}).

The average values of the corrections $\left(\tilde{\gamma}_0/\gamma_0\right)^\beta$ ($\beta = 0.925$) and $\tilde{\beta}_0/\beta_0$ calculated in bins of shower energy for the hybrid data sample selected to evaluate $E_{\rm inv}$ are shown in the right panel of Fig.~\ref{fig:EinvCorr}. Note that $\tilde{\gamma}_0$ and $\gamma_0$ refer to data, and \qgsjet II-03 simulations for the mixed composition, respectively. Here $\tilde{\beta}_0/\beta_0$ represents the muon number excess obtained in inclined events with respect to the muon number predicted by \qgsjet II-03~\cite{QGSJet2} for a mixed composition of 50\% protons and 50\% iron. It is estimated from the parametrization of $N_{19}$  as a function of energy~\cite{Spectrum-ICRC2017} and the parametrizations of $R_{\mu}$ for the  \qgsjet II-03 simulations for proton and iron primaries~\cite{AugerMuonSize}:
\begin{equation}
\frac{\tilde{\beta}_0}{\beta_0} = N_{19}(E) ~ \frac{ R_{\mu}^{\rm p} \left( 10^{19}~{\rm eV} \right) }{ \left(   R_{\mu}^{\rm p}(E) + R_{\mu}^{\rm Fe}(E) \right)/2 }  ~.
\label{eq:N19Corr}
\end{equation}

\section{$\einv$ parametrization as a function of $E_{\rm cal}$ including the extrapolation to low energies and the zenith angle dependence}
\label{Appendix_B}

The function that parametrises $\einv$ as a function of $\ecal$ and extrapolates to $10^{17}$ eV is
\begin{eqnarray*}
E_{\rm inv} & = & f(\theta) ~E_{\rm inv}^{l}  + f(\theta)~\frac{1}{2}~\left[  1 + \tanh\left(  K ~\log_{10}\frac{\ecal}{E_{\rm cal}^A}  \right)  \right]  \\
               &     &~\left(  E_{\rm inv}^{h}  -  E_{\rm inv}^{l}  \right) \ ,
\end{eqnarray*}
where $E_{\rm inv}^{h}$ is the parametrization obtained from the analysis of the inclined events (see Sec.~\ref{Sec:Einv_data}) and $E_{\rm inv}^{l}$ describes the extrapolation down to low energies:
\begin{eqnarray*}
E_{\rm inv}^{l} & = & a ~\left( \frac{ E_{\rm cal}^A } {10^{18} ~{\rm eV} } \right)^b   ~ \left(  \frac{E_{\rm cal}}{E_{\rm cal}^A} \right)^{b_{\rm extr}}     \\
E_{\rm inv}^{h} & = & a ~\left( \frac{ E_{\rm cal} } {10^{18} ~{\rm eV} } \right)^b    \ .
\end{eqnarray*}

The hyperbolic tangent ensures a smooth transition between the functions $E_{\rm inv}^{l}$ and $E_{\rm inv}^{h}$ that describe the invisible energy below and above $E_{\rm cal}^A$, respectively, $E_{\rm cal}^A$ being the energy at which Auger measures a break in the elongation rate~\cite{Auger-mass}.
\begin{table}[h]
\caption{Values of the parameters needed to calculate the invisible energy parametrization.}
\label{tab:Einv-Ecal-par-extr}
\begin{center}
\renewcommand{\arraystretch}{1.5}
\begin{tabular}{c c |c c c| c c c c}
$a\left[ 10^{18} {\rm eV} \right]$    &    $b$     &     $E_{\rm cal}^A\left[ 10^{18} {\rm eV} \right]$     &  $K$    &     $b_{\rm extr}$     & $f_1$  & $f_2$  & $f_3$  \\ \hline
 0.179  &   0.947 & 1.95 & 4 & 0.846  & -0.265  & 0.489 & -0.441 \\
\end{tabular}
\end{center}
\end{table}

The function $f(\theta)$ describes the zenith angle dependence of $\einv$. It is parametrized with a third degree polynomial that is normalised to 1 at the average zenith angle of the inclined hybrid events ($66^\circ$):
\begin{eqnarray*}
f(\theta) & = & 1 + f_1 \left(\cos\theta - \cos 66^\circ \right) + f_2 \left(\cos\theta - \cos 66^\circ \right)^2 \\
 & & +~  f_3 \left(\cos\theta - \cos 66^\circ \right)^3 .
\end{eqnarray*}

The values of the parameters defining the invisible energy parametrization are shown in table~\ref{tab:Einv-Ecal-par-extr}.

The systematic uncertainty on $\einv$ is parametrized with the following function
\begin{eqnarray*}
\frac{\Delta E_{\rm inv}}{E_{\rm inv}} & = & ~~ \sqrt{   0.12^2 +   \left[  0.15 ~ \left( 1 - \frac{\log_{10} \left(E_{\rm cal}/10^{17}\right) }{\log_{10} \left( E_{\rm cal}^A /10^{17} \right) } \right) \right]^2 }
\end{eqnarray*}
for  $10^{17} ~{\rm eV} < E_{\rm cal} < E_{\rm cal}^A$ and is fixed to $\frac{\Delta E_{\rm inv}}{E_{\rm inv}}=0.12$ for $ E_{\rm cal} > E_{\rm cal}^A $.

\end{appendices}

\end{document}